\documentclass[a4paper,fleqn,usenatbib]{mnras}

\usepackage[T1]{fontenc}
\usepackage{ae,aecompl}
\usepackage{adjustbox}

\usepackage{graphicx}	% Including figure files
\usepackage{amsmath}	% Advanced maths commands
\usepackage{amssymb}	% Extra maths symbols
\usepackage{lscape}
\usepackage{pdflscape}
\usepackage[flushleft]{threeparttable}
\title[Double-peaked Narrow Emission-line Galaxies]{Double-peaked Narrow Emission-line Galaxies in LAMOST Survey}

\author[Mengxin Wang et al.]{
M.-X. Wang,$^{1,2}$
A.-L. Luo,$^{1,2}$\thanks{E-mail: lal@nao.cas.cn}
Y.-H. Song,$^{1}$
S.-Y. Shen,$^{3}$
S. Feng,$^{2,3}$
L.-L. Wang,$^{1,2,4}$
\newauthor 
Y.-F. Wang,$^{1}$
Y.-B. Li,$^{1}$ 
B. Du,$^{1,2}$
W. Hou,$^{1}$
Y.-X. Guo,$^{1,2}$
X. Kong,$^{1,2}$
J.-N. Zhang$^{1}$
\\
$^{1}$National Astronomical Observatories of China, CAS, 20A Datun Road Chaoyang District, Beijing 100012, China\\
$^{2}$University of Chinese Academy of Sciences, Beijing 100049, China\\
$^{3}$Key Laboratory for Research in Galaxies and Cosmology, Shanghai Astronomical Observatory, CAS, 80 Nandan Road, Shanghai 200030, China\\
$^{4}$Information Management, Dezhou University, Dezhou 253023, China
}

\date{Accepted XXX. Received YYY; in original form ZZZ}

\pubyear{2018}

\begin{document}
\label{firstpage}
\pagerange{\pageref{firstpage}--\pageref{lastpage}}
\maketitle

\begin{abstract}
We outline a full-scale search for galaxies exhibiting double-peaked profiles of prominent narrow emission lines, motivated by the prospect of finding objects
 related to merging galaxies, and even dual active galactic nuclei candidates as by-product, from the Large Sky Area Multi-object Fiber Spectroscopic 
 Telescope (LAMOST) Data Release 4. We assemble a large sample of 325 candidates with double-peaked or strong asymmetric narrow emission lines, 
 with 33 objects therein appearing optically resolved dual-cored structures, close companions or signs of recent interaction on the Sloan Digital Sky 
 Survey images. A candidate from LAMOST (J074810.95+281349.2) is also stressed here based on the kinematic and spatial decompositions of the 
 double-peaked narrow emission line target, with analysis from the cross-referenced Mapping Nearby Galaxies at the Apache Point Observatory
  (MaNGA) survey datacube. MaNGA enables us to constrain the origin of double peaks for these sources, and with the IFU data we infer that the 
  most promising origin of double-peaked profiles for LAMOST J074810.95+281349.2 is the `Rotation Dominated + Disturbance' structure. 
\end{abstract}

\begin{keywords}
 galaxies: active -- galaxies: interactions -- galaxies: nuclei
\end{keywords}
%%%%%%%%%%%%%%%%%%%%%%%%%%%%%%%%%%%%%%%%%%%%%%%%%%

\section{INTRODUCTION}

In the hierarchical $\Lambda$ cold dark matter cosmology, galaxy mergers represent a cornerstone stage in the structure formation of galaxies, 
since most bulge-dominated galaxies are believed to harbour a central supermassive black hole (SMBH), binary SMBHs would be the inevitable
 outcome of the current $\Lambda$ CDM cosmological paradigm. Discoveries of merging galaxies or dual active galactic nuclei (AGNs) are 
 helpful both for understanding galaxy evolution and for exploring fundamental physics. As the frequency of occurrence and the statistical 
 properties of dual AGNs would provide the helpful insights into the hierarchical merger paradigm of galaxy evolution and the influence of 
 mergers on AGN fueling \citep[e.g.][]{2011MNRAS.418.2043E,2012ApJ...745...94L}. The merger of massive galaxies can also invoke a 
 series of phenomena which are rather different from those expected from single SMBH, such as the formation of large elliptical galaxies 
 \citep{2007MNRAS.377..957H}, stimulation of quasars' activity and the SMBH growth \citep{2005ApJ...625L..71H}, and the possibly 
 observable strong bursts of gravitational radiation \citep{2007MNRAS.377..957H}.
 
The search for merging galaxies and AGNs has therefore received great attention, method utilizing the double-peaked narrow emission 
lines regions of object spectrum has been explored in candidates search. As we know, if two galaxies merger to kpc scale, and the 
adjacent narrow line regions (NLRs) are sensed by a single spectrograph slit or fiber, double-peaked emission lines would appear in 
the spectra for two cores, thus providing the potential for finding binary or dual AGN candidates at intermediate separation that can not be spatially resolved. 
AGNs with double-peaked narrow emission-line features have been discovered 
since the 1980s \citep[e.g.][] {1981ApJ...247..403H,1984ApJ...281..525H,1985Natur.318...43K}, and were mainly 
associated with outflows or rotating 
NLRs \citep{2005ApJ...627..721G}. Since the suggestion that double-peaked 
emission line profiles could also be produced by merging galaxies and AGNs
 \citep{2009ApJ...705L..76W}, more efforts have been invested into the
search for related objects and study of the statistical properties of 
double-peaked emission line candidates on the basis of several fundamental 
issues exhibiting an influence on the evolution of galaxies and AGNs 
\citep[e.g.][]{2009ApJ...705L..76W,2010ApJ...708..427L,2010ApJ...716..866S,2012ApJS..201...31G}.
 To date, there are no more than 30 resolved dual AGNs reported 
 \citep[e.g.][]{2014Natur.511...57D,2015ApJ...813..103M}, which should be 
 confirmed through high-resolution imaging by X-ray or radio observations. 
 Some convincing examples were discovered serendipitously, such as OJ 287 \citep{1988ApJ...325..628S,1996ApJ...460..207L},
  LBQS 0103-2753 \citep{2001ApJ...549L.155J}, NGC 6240 \citep{2003ApJ...582L..15K}, 3C 75 \citep{2006A&A...453..433H}, 
  0402+379 \citep{2006ApJ...646...49R}, Mrk 463 \citep{2008MNRAS.386..105B}, COSMOS J100043+020637 
  \citep{2009ApJ...702L..82C}, NGC3393 \citep{2011Natur.477..431F}, NGC 326 \citep{2012ApJ...746..167H}, PG 1302-102 
  \citep{2015Natur.518...74G}. Additional dual AGNs were confirmed with the help of detecting parent samples with double-peaked 
  narrow emission lines. \citet{2013ApJ...762..110L} confirmed the binary-AGN scenario for two targets with high-resolution optical 
  and X-Ray imaging from a parent sample of Type 2 AGNs with double-peaked narrow emission lines. Based on a parent sample 
  of 340 unique AGNs with double-peaked narrow emission lines identified in Sloan Digital Sky Survey (SDSS) at 0.01 < z < 0.7, 
  \citet{2015ApJ...806..219C} identified six dual AGNs and dual/offset AGNs using X-rays, including an extremely minor merger (460:1)
   that may indicate a dwarf galaxy hosting an intermediate mass black hole. \citet{2015ApJ...813..103M} also confirmed the presence 
   of dual AGNs in three galaxies using the optical long-slit spectroscopy and high-resolution multi-band Very Large Array (VLA) observations.

As it has been reported, other processes not related to merging galaxies or AGNs can also potentially explain the phenomenon of narrow
 double-peaked emitting, including disturbed NLRs involving biconical outflows
 \citep[e.g.][]{1994AAS...185.2705G,2011ApJ...727...71F,2011ApJ...735...48S,2012ApJ...745...67F}, the local interaction of radio jets 
 with NLR clouds \citep[e.g.][]{2010ApJ...716..131R}, rotation-dominated regions with disturbance or obscuration 
 \citep[e.g.][]{2016ApJ...832...67N,2015ApJ...813..103M}, special NLR geometry, or a single AGN illuminating the interstellar 
 media of both galaxies \citep[see][]{2009ApJ...705L..20X}. Besides the dual AGN candidates that can be expected, the active 
 galaxies that own NLRs with disturbed kinematics, which manifest as double-peaked emission lines profiles, can also enable us to 
 probe the origin of disturbed NLRs and indicate complex dynamical processes related to the AGN feeding and feedback 
 \citep{2016ApJ...832...67N,2015ApJ...813..103M}. \citet{2015ApJ...813..103M} has critically investigated the nature of the ionizing 
 sources of 18 optically identified double-peaked AGNs from SDSS, and pinpointed the origin of the double-peaked emission 
 features for these samples based on optical long-slit spectroscopy and high-resolution ($\sim$ 0.2 arcsec)  VLA multi-band 
 observations in the range of 8-12 GHz. Among these, only $\sim$ 15\% of the sources are produced by dual AGNs, and gas 
 kinematics would account for $\sim$ 75\% of the double-peaked narrow emission lines, while the remaining 10\% are ambiguous 
 cases. Furthermore, \citet{2016ApJ...832...67N} have created a classification criteria for kinematic origin of double-peaked narrow 
 emission line targets based on optical long-slit spectroscopy of a complete sample 71 double-peaked AGNs at z < 0.1 in SDSS.

Despite the low dual AGNs confirmation rate from follow-up spatially observations of the reported double-peaked emission-line 
candidates \citep[e.g.][]{2011ApJ...735...48S,2012ApJ...753...42C}, the confirmed systems and investigation of kinematic nature 
from double-peaked narrow emission line candidates still lend strong support to the implications and significance of systematically 
searches for double-peaked narrow emission line objects\citep[e.g.][]{2013MNRAS.429.2594B,2013ApJ...762..110L,2015ApJ...806..219C}.

In this paper, we carry out a full-scale search in the LAMOST DR4 database for galaxies and AGNs with double-peaked 
emission-line features exhibited by typical emission lines, including H$\beta$, [O~{\sc iii}], H$\alpha$ and [N~{\sc ii}]. With 
reasonable criteria for fluxes, velocity shifts, width and intensity ratios of narrow lines and careful visual inspection, we establish a 
large sample of 325 candidates. In Section 2, the LAMOST spectroscopic observation is introduced and the selection procedures, 
adopted criteria and final samples are described. In addition, we critically inspect SDSS images for the selected targets and display 
interesting results in Section 3. A cross-referenced target with MaNGA and relevant analysis on its origin of double peaks feature is stressed in Section 4, followed by a
 brief summary in Section 5. We adopt a standard cosmology of $\Omega_{\rm M}$=0.3, $H_0$=70 km s$^{-1}$Mpc$^{-1}$, and 
 $\Omega_{\Lambda}$=0.7 throughout the work. All quoted wavelengths are presented in vacuum units. 
 
 \section{OBSERVATION AND SELECTION}

\subsection{LAMOST Survey}
\label{sec:maths} % used for referring to this section from elsewhere

LAMOST (also called the Guo Shou Jing Telescope) is a 4 meter quasi-meridian reflecting Schmidt telescope, with 4000 fibers 
mounted on its focal plane, exhibiting high spectral acquisition rate \citep[see][]{2012RAA....12.1197C}. Its spectral range extends 
from 3700 to 9000 {\AA} with a resolution of {\it R} $\sim$ 1800. After a two-year commissioning phase and one-year pilot survey, 
the LAMOST regular surveys were initiated in Oct 2012. It focuses mainly on Galactic stars survey, but also observes a significant 
fraction of extra-galactic targets \citep{2015RAA....15.1089L}, which consist of two main parts: a galaxy survey and a QSO survey. 
For galaxies, the surveying area is split into two regions (to avoid Galactic plane). One is in the Northern Galactic Cap (NGC) region, 
which shares footprint with the SDSS Legacy Survey, aiming to observe the SDSS main sample galaxies (with {\it r} <17.75), which 
are failed to be spectroscopically observed due to SDSS fiber collision \citep{2002AJ....123..485S}, such ``missed" targets are 
referred to as complementary galaxy samples in the LAMOST survey. The other region of the LAMOST galaxy survey is the 
Southern Galactic Cap (SGC), the team has a long-range goal of galaxy surveying in the LAMOST SGC 
footprint (with {\it b} < - $30^{\circ}$ and $\delta$ > - $10^{\circ}$), aiming to observe galaxies with with {\it r} <18, and a subset of 
blue galaxies down to {\it r} <18.8. Apart from these two well-defined parts, some bright-infrared galaxies are also selected as extra
 surveying targets which are chosen from infrared surveys such as WISE, IRAS and with {\it Herschel}. The LAMOST DR4, 
 based on the pilot survey and the regular survey from OCT 2011 to JUNE 2016, has collected $\sim$ 117000 galaxy and $\sim$ 
 36000 QSO spectra.

\subsection{Sample and Method }
\label{sec:Sample_and_Method} % used for referring to this section from elsewhere

Our parent sample is the LAMOST DR4 galaxy and QSO dataset, which includes objects (153348 in total) spectrally 
classified as galaxies or QSOs by the LAMOST pipeline. The spectra with redshift larger than 0.8 are removed from our sample,
guaranteeing that the prominent [O~{\sc iii}]$\lambda$5007 emission lines are included in the wavelength coverage. As for 
LAMOST data product, the signal-to-noise (S/N) is defined by using the concept of inverse-variance, the mean S/N in a wavelength band 
can be averaged from the S/N of each pixel in the band range, and mainly represents the uncertainty of spectra continuum, see 
more in \citet{2015RAA....15.1095L}. Since it is possible to show strong emission lines despite a noisy spectrum, we impose a 
loose signal-to-noise (S/N) cutoff (S/N > 2) to avoid missing noisy but interesting objects. In light of the uncertainty of EW 
calculation brought about by measurement errors, noise and inaccurate estimation resulting from uniform-length wavelength-window 
adoption for different emission line widths of spectra with low S/N, the selection criteria for emission-line candidates is temporarily 
more relaxed than it was in previous works \citep[see][]{2014RAA....14.1234S}, the galaxies and QSOs with EW(H$\alpha$) <-3 {\AA} 
and EW([O~{\sc iii}]$\lambda$5007 < 0 {\AA} (here negative equivalent width stands for emission lines) are selected for the subsequent 
steps. The number of spectra satisfying the above criteria amounts to 55082. As an initial screening step, we conduct a thorough and 
strict visual inspection of these remaining spectra, with the purpose of finding candidates displaying well-detected double-peaked or 
strong asymmetric profiles in the region of several key emission lines, the lines considered here are 
H$\beta$-[O~{\sc iii}$\lambda\lambda$4959,5007 and H$\alpha$-[N~{\sc ii}]$\lambda\lambda$6548,6584. As for the H$\alpha$-[N~{\sc ii}] 
set, since their rest-frame wavelengths are proximately located, it is likely that their multi-components may blend with each other and 
would exhibit complex features. It is also requested that the split or asymmetric features show similar trend for all the interesting narrow 
emission lines of each selected spectra. After this narrow filtering, we limit the sample to 1324 spectra.

The LAMOST spectra are taken through fibers with a fixed 3.2 arcsec aperture, large enough to let through the light from nucleus 
and the stellar component from the host galaxy, for instance, a 3 arcsec aperture would subtend about 4 {\it h\/}$^{-1}$ pc for a 
sample at a median redshift (z = 0.1), galaxies with higher redshift would exhibit more host galaxy composition in their observed 
spectra, which need be subtracted for more reliable emission-line analysis. Here, we resort to a stellar population synthesis 
program named STARLIGHT to estimate the underlying continuum. Prior to the STARLIGHT process, we also complete the 
foreground extinction correction using the reddening maps from \citet{1998ApJ...500..525S} and measure the systemic velocity of 
each galaxy for the following operations. 

Most of our objects spectra exhibit stellar absorption features over the observed spectral range, we fit the stellar lines to derive the 
systemic velocity. The template is derived from 100 absorption-lines galaxy spectra from LAMOST, which own high-precision 
redshift (their z offset between LAMOST and cross-referenced SDSS spectra is smaller than $10^{-5}$). We adopt the redshift 
from LAMOST pipeline as the initial guess, and take trial values that are advanced in steps representing the pixels of each 
spectrum. For each trial redshift, we fit the stellar lines with the synthesized template, at the rest-frame spectral region 
5100-5950\AA, containing the MgB and NaD lines, and get their reduced $\chi^2$ value. A $\chi^2$(z) curve can be defined 
in the probed redshift range and the best absorption line redshift is determined by the fit having the minimum $\chi^2$ value. 
We also reject the output for spectra whose $\chi^2$ valley is not well-defined. The uncertainty of the redshift is evaluated by the 
width of the $\chi^2$ minimum at which the $\Delta\chi^2$ equals 1. The host redshifts are derived with an average uncertainty 
of 28 km s$^{-1}$  for sample in the rest frame. For galaxies with failed $\chi^2$ fit, or without clear signs of absorption lines, the 
redshifts and stellar velocity dispersions from LAMOST pipeline are adopted. As LAMOST conducts the spectroscopic survey 
without photometric cameras equipped, it could only offer relative flux for each object, the stellar spectra created by 
STARLIGHT is weighted with a cubic polynomial here (if necessary) to compensate the continuum uncertainty brought by 
existing reddening and flux calibrating discrepancy, without affecting the absorption- and emission-line features. 

Pure emission-line spectra is composed of three kinetic component groups, including the narrow emission lines (H$\beta$, 
[O~{\sc iii}], H$\alpha$ and [N~{\sc ii}]), [O~{\sc iii}] wings and broad Balmer emission lines. We set up three models to 
describe each emission line: single gaussian for only one component, double-gaussians for the blueshifted and redshifted 
systems (here we refer to the emission-line component at lower redshift as ``blueshifted system", and the part at 
higher redshift as ``redshifted system'') and three-gaussians for emission lines comprising two components and a wing ([O~{\sc iii}]) 
or broad balmer component (H$\alpha$, H$\beta$). For each spectrum, we select the fitting regions manually, shielding out the 
side-effects possibly brought by interfering lines nearby and reducing the fitting sensitivity to uncertainties from the continuum. 
As for the multi-gaussians fit, the procedure assumes a consistent velocity width for all three categories, H$\alpha$ and H$\beta$, 
[N~{\sc ii}]$\lambda\lambda$6548,6584 and [O~{\sc iii}]$\lambda\lambda$4959,5007 respectively, based on their different  
excitation mechanisms \citep{2012AJ....144..144B}, and comparable velocity offsets between blueshifted and redshifted 
components for H$\beta$-[O~{\sc iii}]$\lambda\lambda$4959,5007 and H$\alpha$-[N~{\sc ii}]$\lambda\lambda$6548,6584 
lines set, with velocity shifts and widths for [O~{\sc iii}] wings and broad Balmer component being adjustable in the fittings. 
Fluxes for all fitting components are set free, except for the 3 : 1 intensity ratio of [N~{\sc ii}]$\lambda$6584 
to [N~{\sc ii}]$\lambda$6548 and [O~{\sc iii}]$\lambda$5007 to [O~{\sc iii}]$\lambda$4959, required by atomic physics 
\citep[see][]{2006agna.book.....O}, and all FWHMs are corrected for the instrumental resolution. The fit is optimized in a 
non-linear least-squares fashion and the model which has the smallest reduced $\chi^2$ is taken as the final selection. 
We inspect each fit and find 362 objects achieving a good profile fit using the multi-gaussians model in this procedure.

\begin{figure}
\centering
\includegraphics[width=0.49\textwidth]{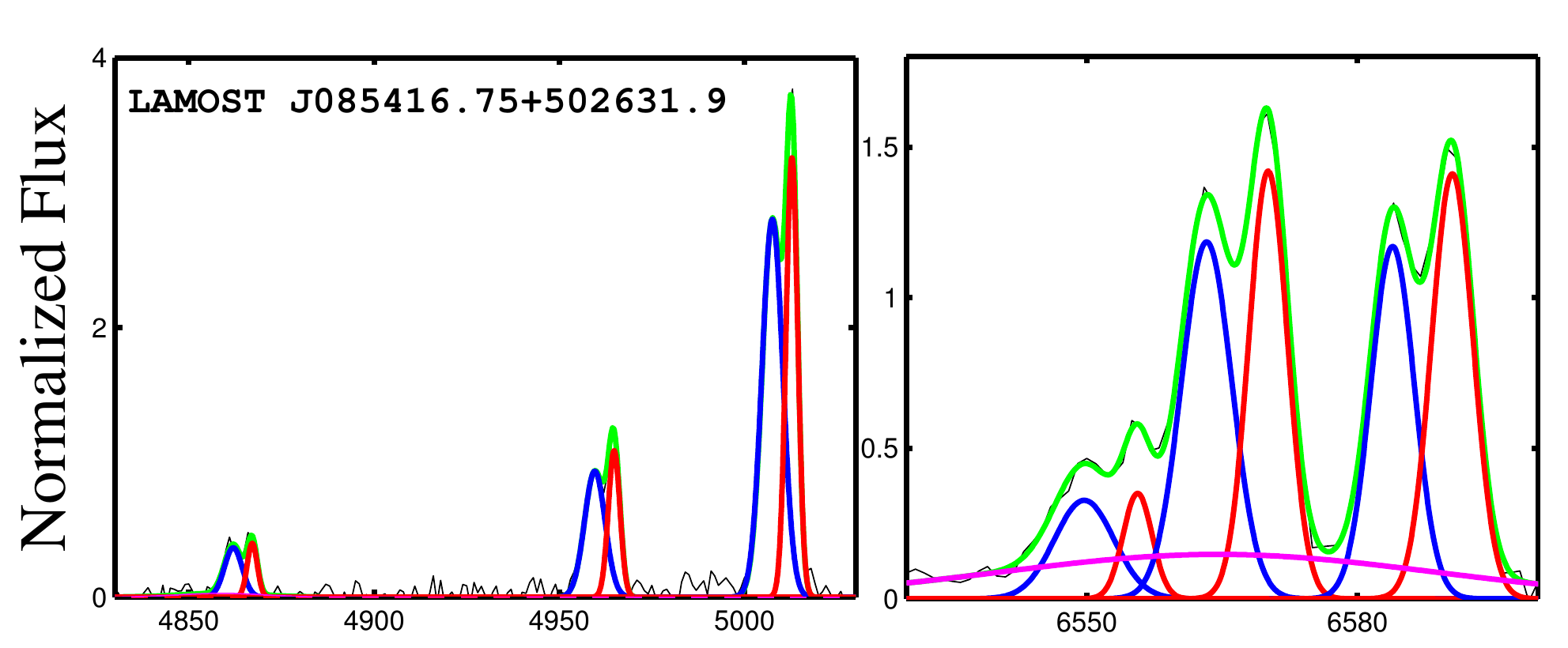}
\includegraphics[width=0.49\textwidth]{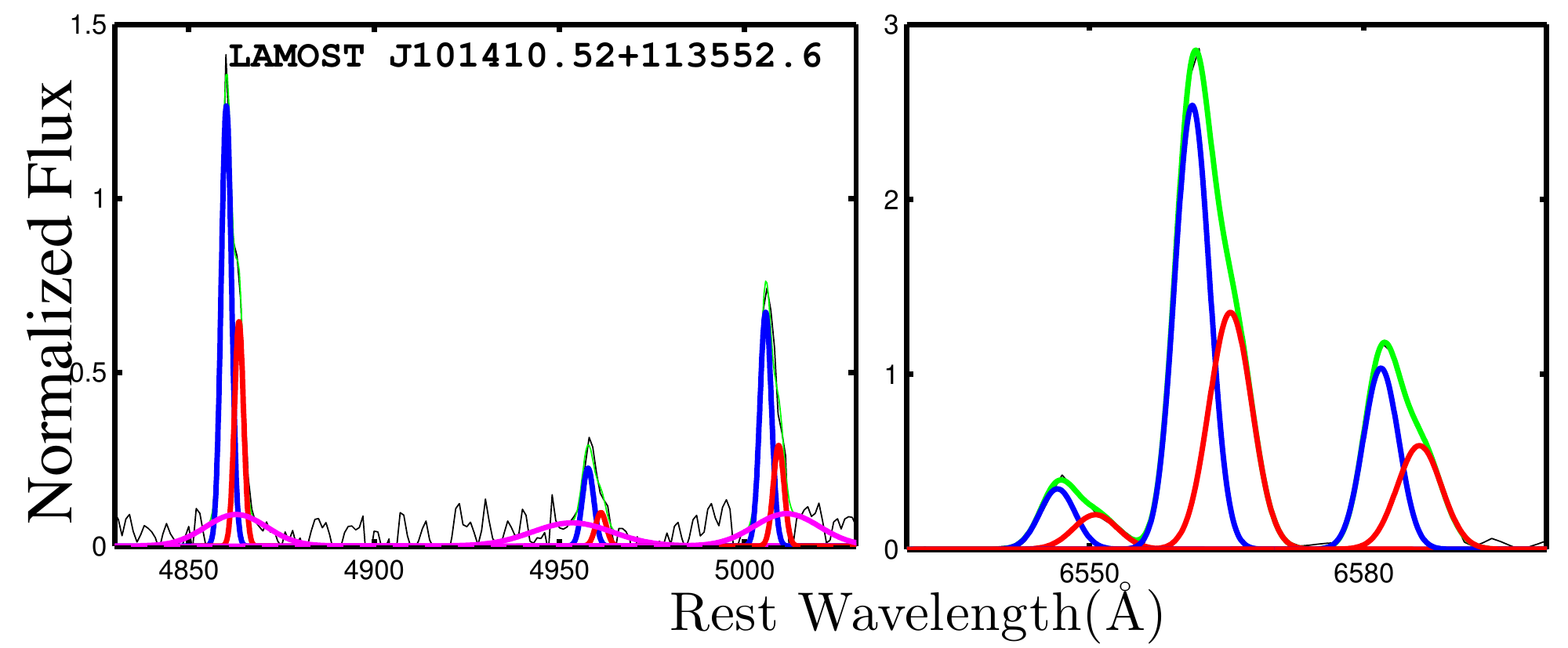}
\caption{Gaussian decomposition for spectra example of double-peaked galaxies and asym narrow emission-lines 
galaxies, respectively. Both panel zoom into the H$\beta$, [O~{\sc iii}] and the H$\alpha$-[N~{\sc ii}] regimes, and 
display the multi-gaussians fit for these narrow emission lines. The upper panel represents a double-peaked sample, 
while the lower panel reveals an example of the asymmetric emission-line galaxies. The blue and red lines in both 
panels represent the blueshifted and redshifted component, respectively, with the carmine lines showing intermediate 
broad components, and the green lines representing the final fitting model.}
\label{examples1}
\end{figure}

The separation between the blueshifted and redshifted system should be large enough to reveal reliable double peaks or strong asymmetric feature. In 
light of the consistent spectral resolution between LAMOST and SDSS spectra, the criteria for selecting candidates in 
\citet{2012ApJS..201...31G} also holds for spectra from LAMOST:
\begin{eqnarray*}
\Delta V/{\rm FWHM_{min}}>0.8,~~{\rm or}~~\Delta V > 200 {\rm km ~s^{-1}},
\end{eqnarray*}
where $\Delta V$ represents the relative velocity gap between the redshifted and blueshifted system,
$\rm FWHM_{\rm min}=\min(\rm FWHM_{\rm r}, FWHM_{\rm b}$), $\rm FWHM_{\rm r}$ and
$\rm FWHM_{\rm b}$ are FWHMs of the two components, respectively. This criteria in Ge's work is defined using 
the Monte Carlo simulations to maximally select potential galaxies with double-peaked profiles, unavoidably covering many asymmetric emission-lines sources. 
After this screening, 325 targets are obtained. Among these, 188 objects exhibit a clear ``trough" in their profiles, 
while the other 137 ones show obvious asymmetric features. Figure \ref{examples1} displays a zoom on the two 
sets of  H$\beta$, [O~{\sc iii}] and H$\alpha$-[N~{\sc ii}] complexes for double-peaked and asymmetric samples, respectively.

There are about four prior works on double-peaked or dual candidates searches within SDSS survey 
\citep[see][]{2009ApJ...705L..76W,2010ApJ...708..427L,2010ApJ...716..866S,2012ApJS..201...31G} and one 
relevant searching work based on LAMOST DR1 dataset \citep[see][]{2014RAA....14.1234S}. Samples in this 
paper has already excluded the overlapping 20 objects presented by Shi. Since it has been reported in Ge's work, 
his catalog covers all the known candidates from the SDSS galaxy samples (DR7), we cross reference our selected 
325 targets with the ones provided in his work, and find 85 objects included, the remaining 240 objects are firstly 
reported as double-peaked or asymmetric narrow-line galaxies here. Within our catalog, 101 targets are spectroscopically 
observed for the first time by LAMOST.

\subsection{BPT Classification}
\label{sec:BPT Classification} % used for referring to this section from elsewhere

\begin{figure}
\centering
\includegraphics[width=0.48\textwidth]{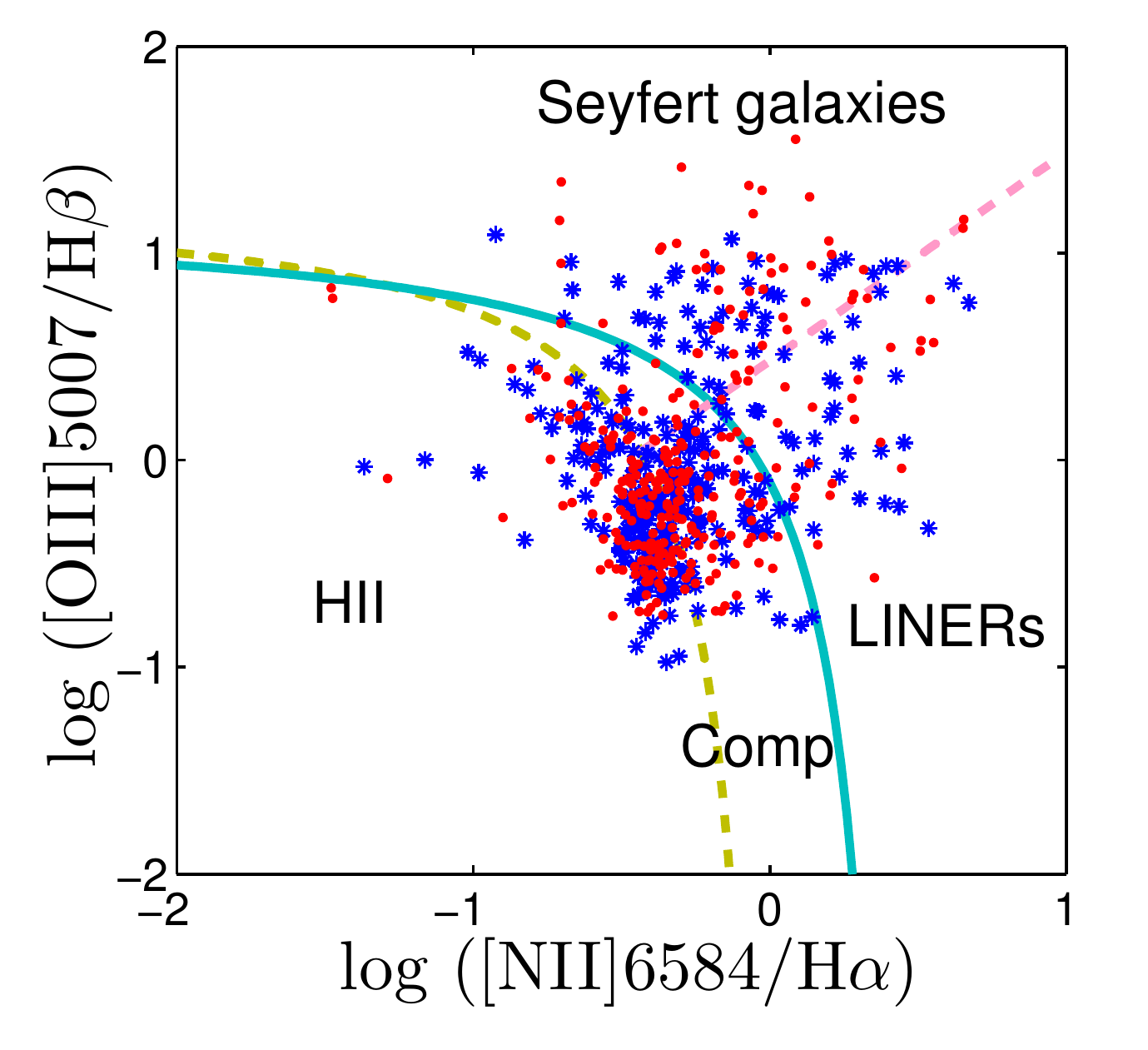}
\caption{BPT diagram of blueshifted and redshifted components of the selected sample. The green solid and yellow dotted lines represent the classification lines suggested by \citet{2001ApJ...556..121K}, and \citet{2003MNRAS.346.1055K}, respectively, and the pink dotted lines represents the alternative borderline suggested by \citet{2010MNRAS.403.1036C} for Seyfert/LINER classification. The blue cross symbol represents the blueshifted component, and the red dot symbol corresponds to the redshifted component, of each selected target.}
\label{325bpt}
\end{figure}

As the blueshifted and redshifted components originate from different ionization mechanisms, the nature of each 
component is diagnosed by examining its location in the Baldwin-Phillips-Terlevich (BPT) diagram 
\citep {1981PASP...93....5B}, which can be divided into star-forming (SF) galaxies, composite galaxies, 
LINERs and Seyferts. Here we assume that if a galaxy displaying prominent broad Balmer lines (H$\alpha$ and H$\beta$), 
and at least one of its narrow emission-line components falls into the Seyfert category after subtracting the broad 
component, it would harbour at least one broad-line AGN composition. The FWHM criteria from \citet {2005AJ....129.1783H} 
is chosen to select the broad-line AGN.
\begin{eqnarray*}
{\rm FWHM(H\alpha) > 1200 km~s^{-1}~~ and}~~ h_{\rm H\alpha, broad}/\bar{h}_{\rm H\alpha, narrow}>0.1,
\end{eqnarray*}
or
\begin{eqnarray*}
{\rm FWHM(H\alpha) > 2200 km~s^{-1}},
\end{eqnarray*}
where $\bar{h}_{\rm H\alpha, narrow}=(h_{\rm H\alpha, red}+h_{\rm H\alpha,blue})/2$ is the mean height calculated from the flux of blueshifted and redshifted narrow components of H$\alpha$, and FWHM(H$\alpha$) stands for the FWHM of the broad H$\alpha$ lines. There are 18 objects in the presented sample are identified as the broad-line AGNs. In figure \ref{325bpt} we show the sample in the BPT diagram (with the broad-line AGNs selected above are excluded), with the redshifted and blueshifted components are illustrated by different colors and symbols respectively, the AGN and star-forming (SF) galaxy dividing lines developed by \citet{2003MNRAS.346.1055K} and \citet{2001ApJ...556..121K} are plotted as well. The AGN component selected under Kauffmann's law would contain these AGN and starburst composite galaxies, which take a large share of the sample, while the AGNs created from Kewley's criteria would include only the unambiguously AGN-dominated objects, here we employ Kewley's law for samples' sort and subsequent statistical analysis. Galaxies with relevant weak emission lines (i.e., emission lines with their amplitudes detected with less than 3$\sigma$ significance) and those with redshifts large than 0.39 (H$\alpha$ and [N~{\sc ii}] lie beyond the coverage of LAMOST spectra) are omitted in the diagram and are defined as type `unknown', the amount of these unknown spectra is 11. To distinguish the Seyferts from LINERs, we employ an alternative dividing line derived from \citet{2010MNRAS.403.1036C}, which proposes a more economical and simpler Seyfert/LINER division diagnosis than \citet{2006MNRAS.372..961K}, without the use of extra emission lines and thus avoiding the loss of information for more emission-line samples. The border is described by the equation below, and it does a good job in translating the \citet{2006MNRAS.372..961K} Seyfert/LINER classification to the BPT diagram.
\begin{eqnarray*}
    \rm log_{10}([OIII]/H\beta)=1.01*log_{10}([NII]/H\alpha)+0.48.
\end{eqnarray*}
As the blueshifted and redshifted components for each double peaks could be driven by different mechanism, there exist several grouping combinations for these candidates. In our sample, there are 28 2-Seyfert II AGNs, 10 2-LINERs, 18 LINER+Seyfert II objects, 204 2-SF ones, 13 SF+Seyfert II objects and 23 SF+LINER ones. In summary, this search finds 325 double-peaked narrow emission-line galaxies, and therein 28 dual AGN candidates and 49 offset AGN candidates (one AGN in a merger system)\citep[see][]{2014ApJ...789..112C}. Table \ref{sample_table} lists the velocity offsets, the FWHMs of the two components, also the observation report, and BPT type for the sample, it is organized according to classification type. We compare the fraction of dual/offset AGN candidates in our sample of 325 galaxies with the corresponding fraction in \citet{2012ApJS..201...31G}. These authors classify Seyfert II galaxies and LINRS Type II AGNs, while in our work, we only refer to the ones with Seyfert II feature as AGN candidates, we use the same classification as it is mentioned in his work, and find that the dual/offset AGN candidates' fraction in our sample is $\sim$ 35\%, which is approximately equal to the value $\sim$ 36\% in his work.

\section{SDSS IMAGES OF THE SAMPLE}

\begin{figure*}
\centering
\includegraphics[width=0.15\textwidth]{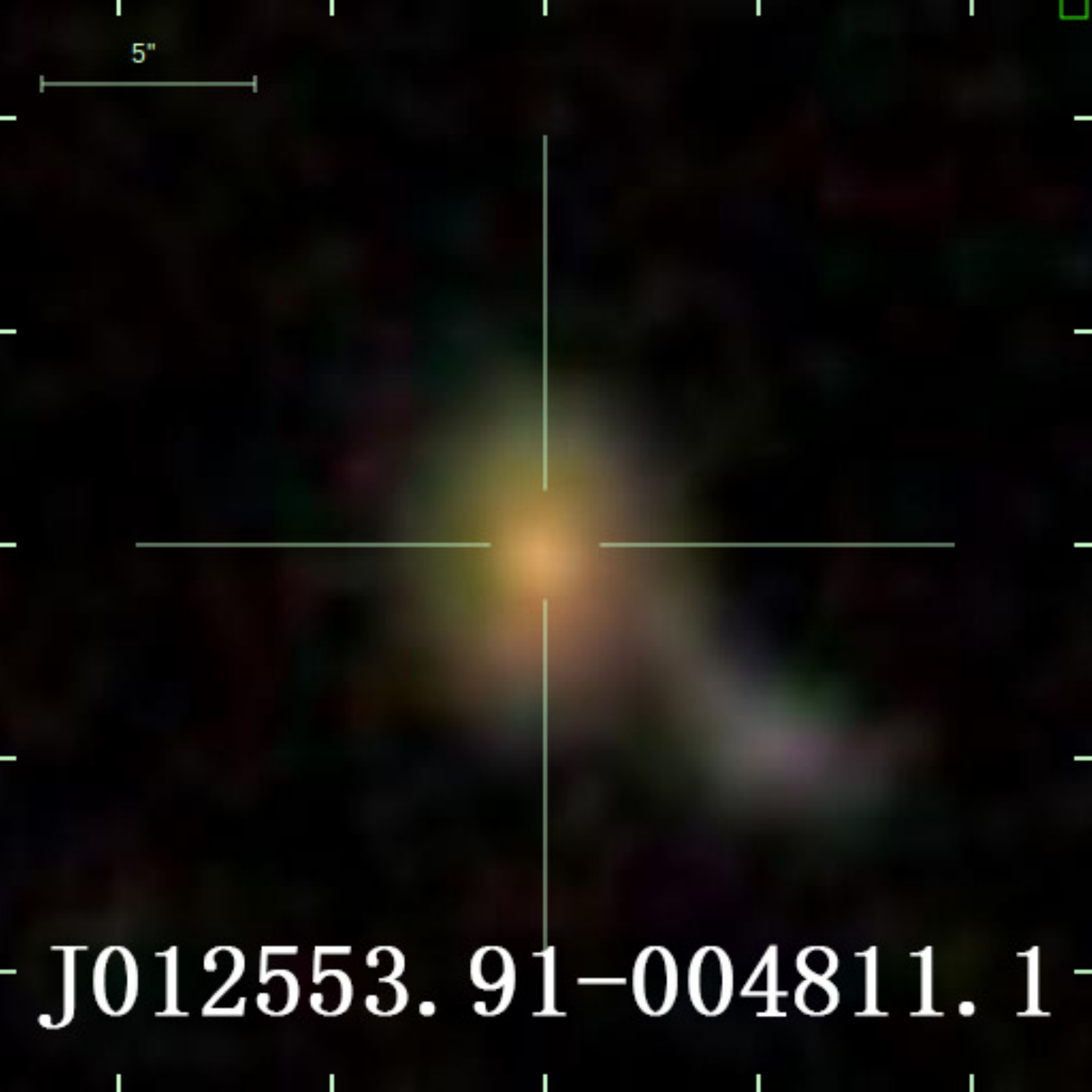}
\includegraphics[width=0.15\textwidth]{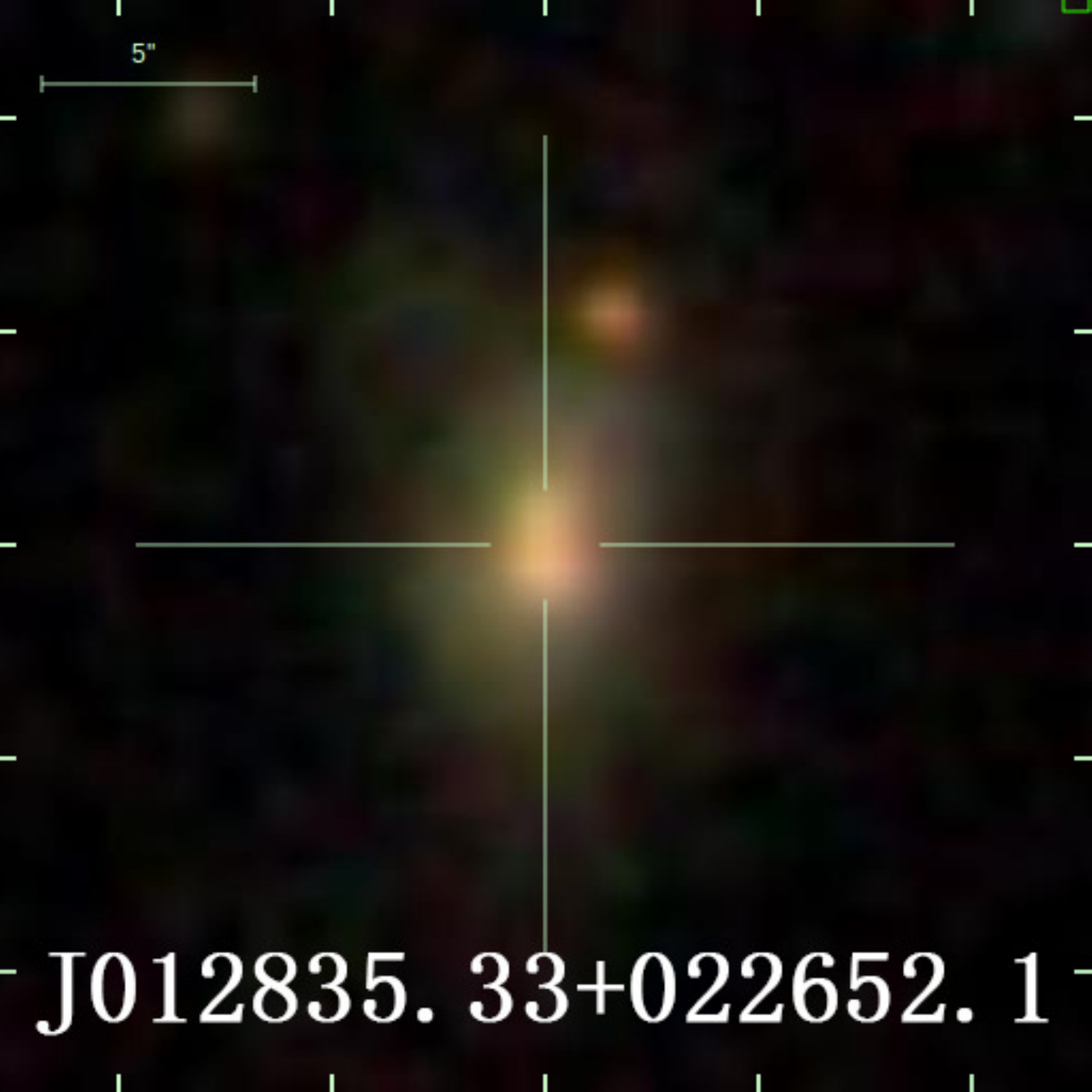}
\includegraphics[width=0.15\textwidth]{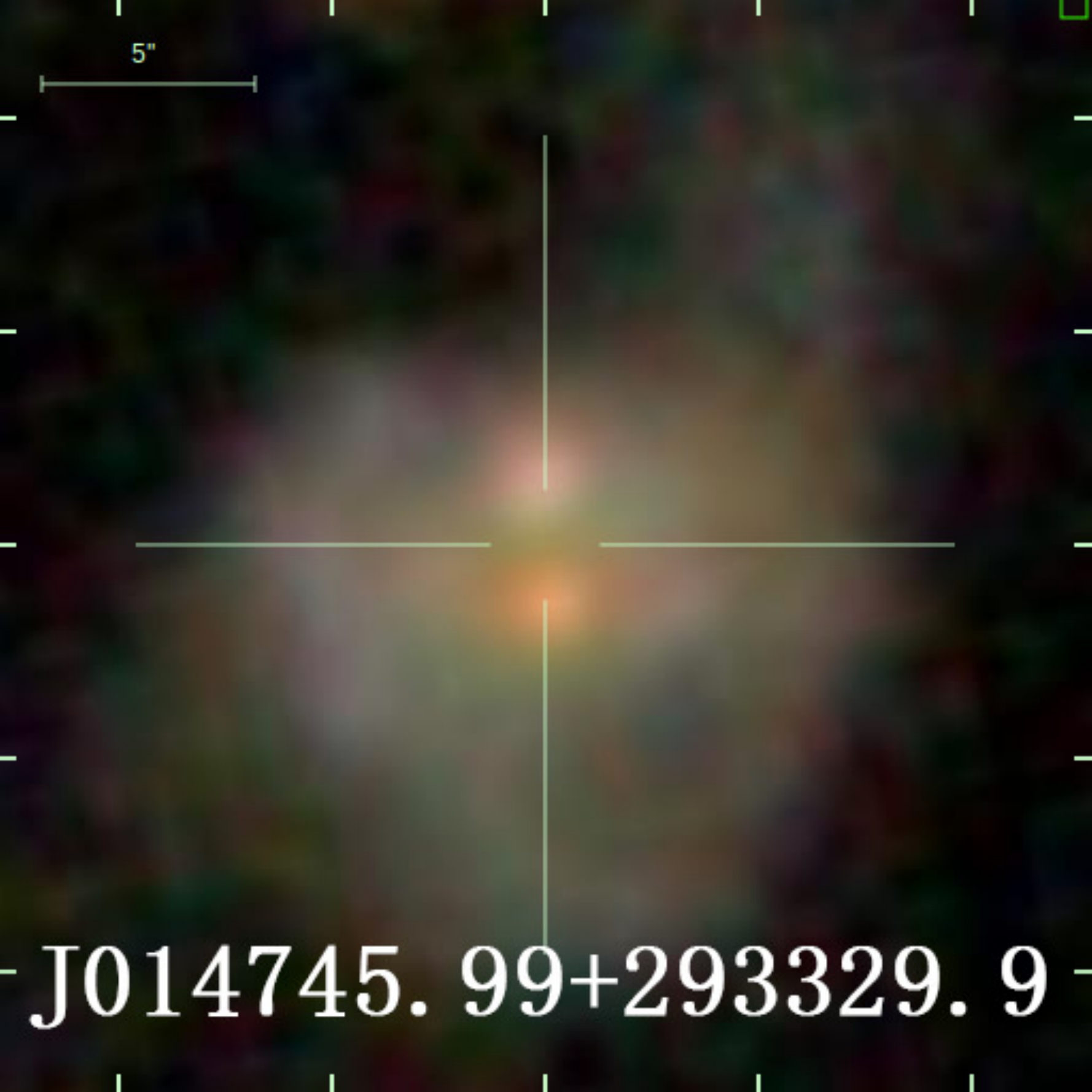}
\includegraphics[width=0.15\textwidth]{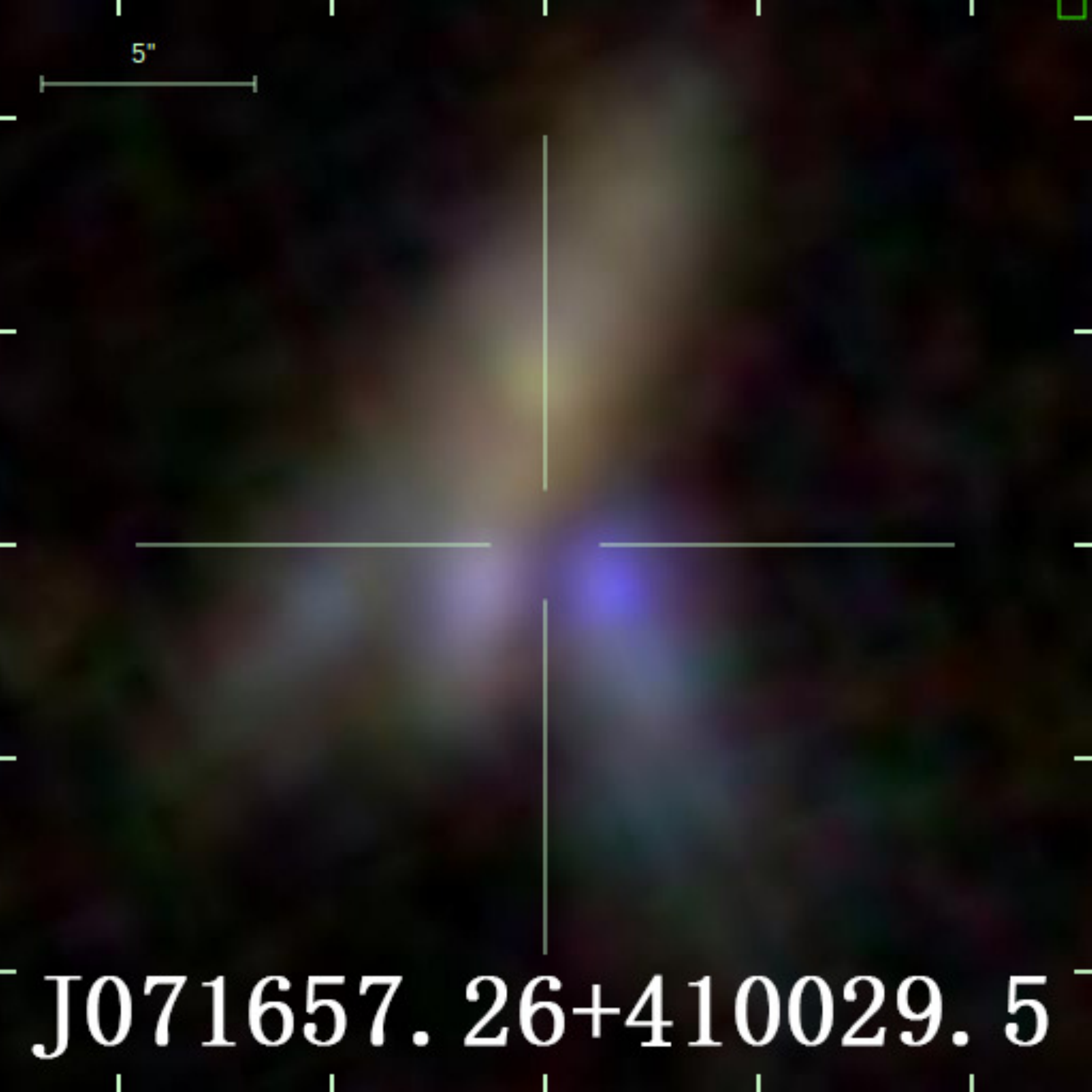}
\includegraphics[width=0.15\textwidth]{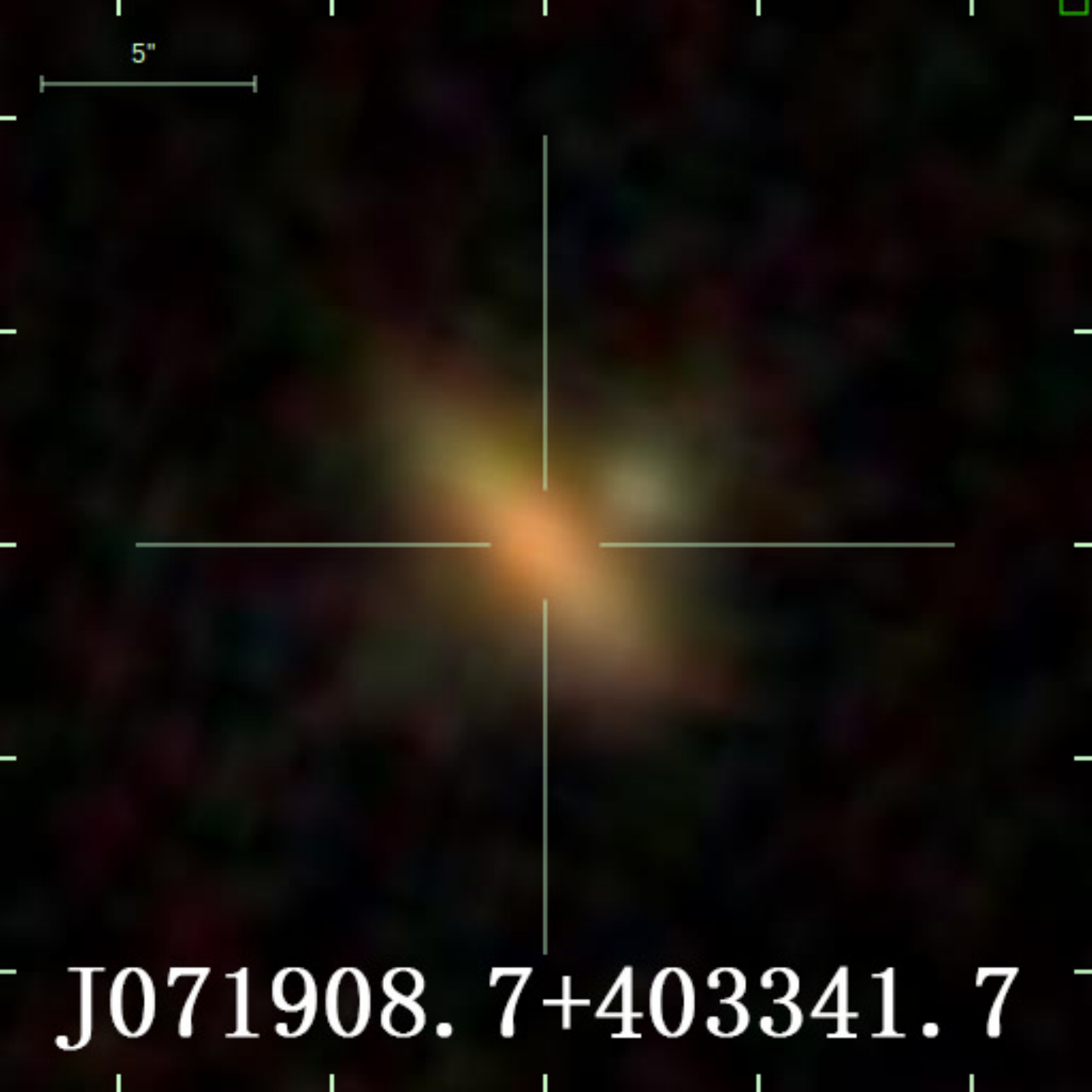}
\includegraphics[width=0.15\textwidth]{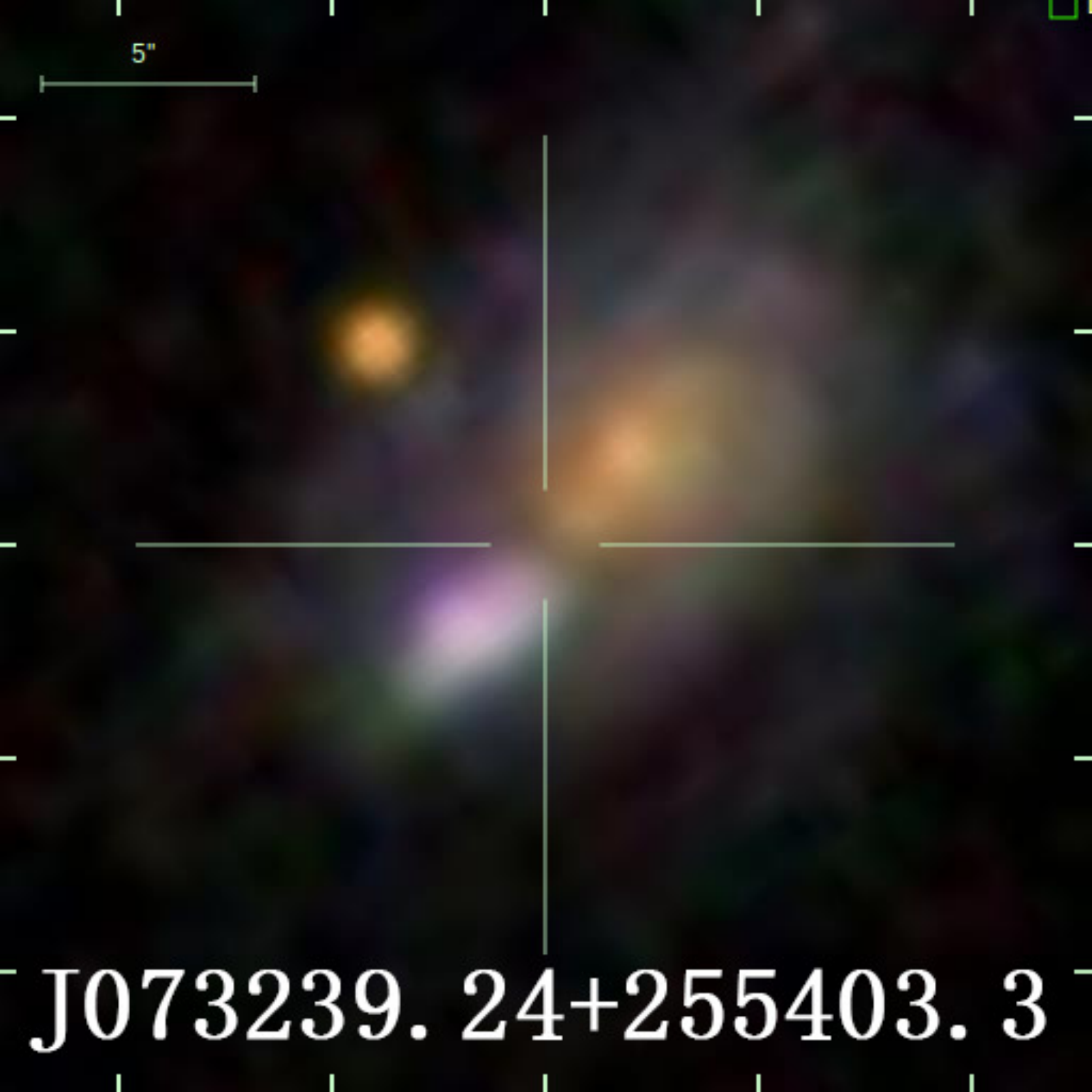}
\includegraphics[width=0.15\textwidth]{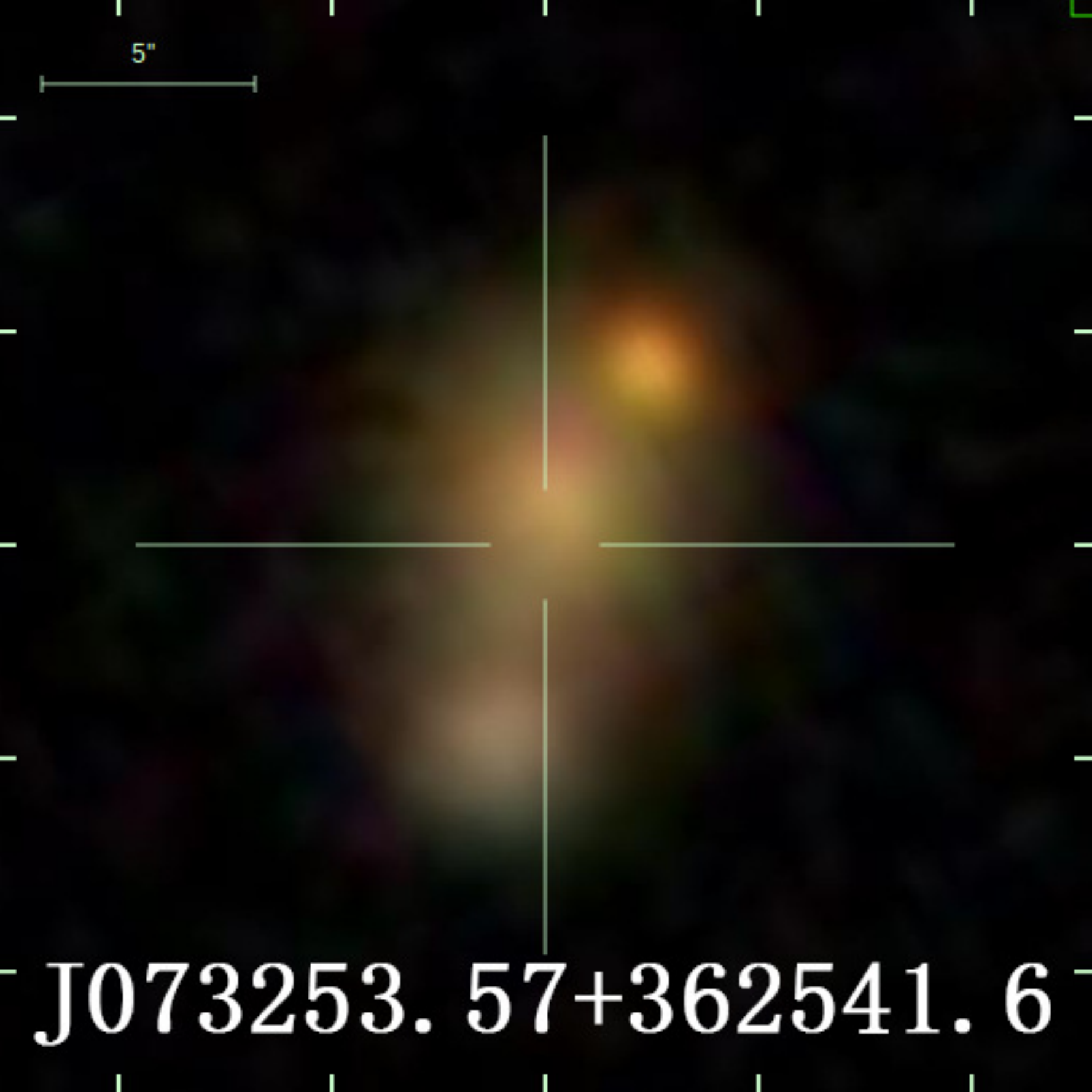}
\includegraphics[width=0.15\textwidth]{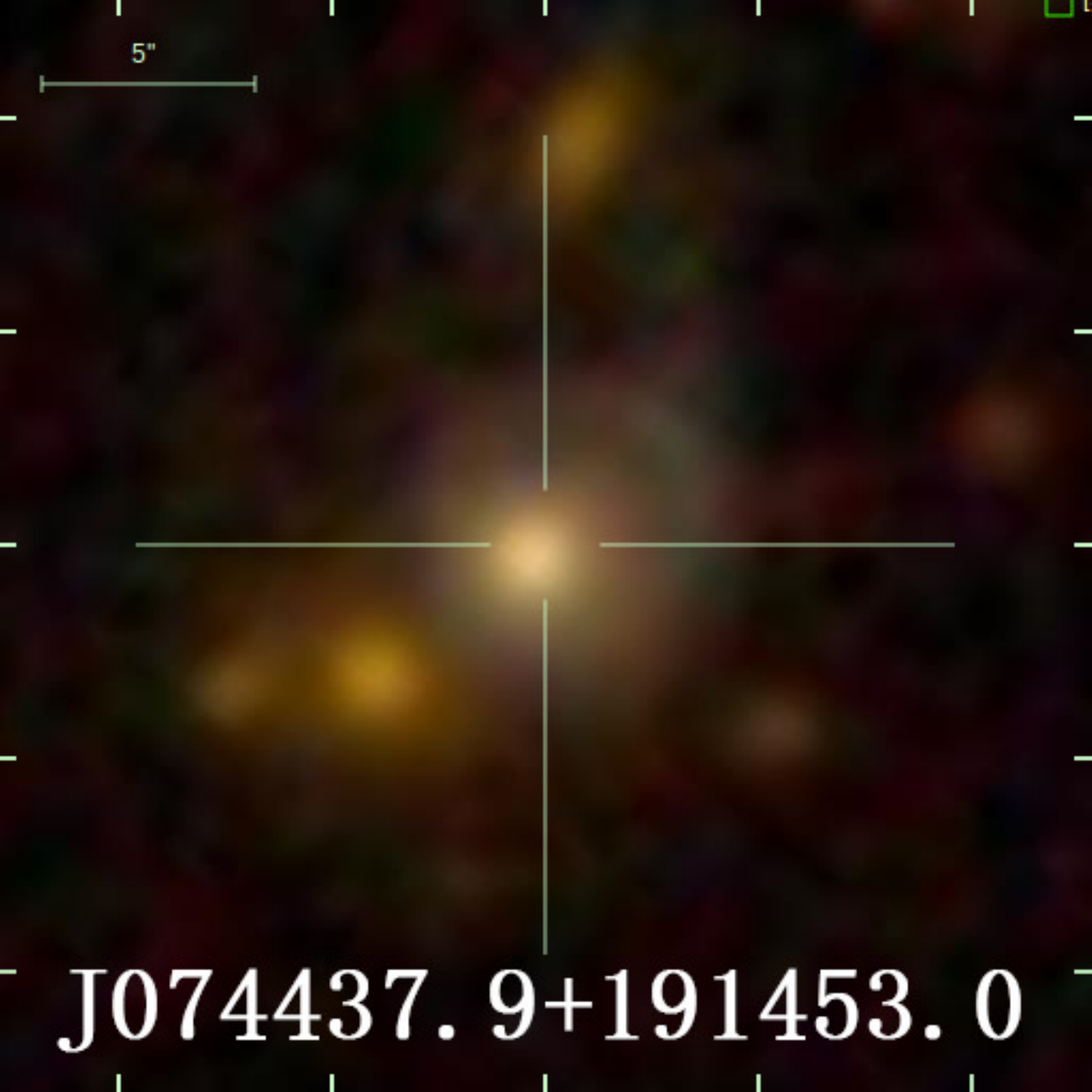}
\includegraphics[width=0.15\textwidth]{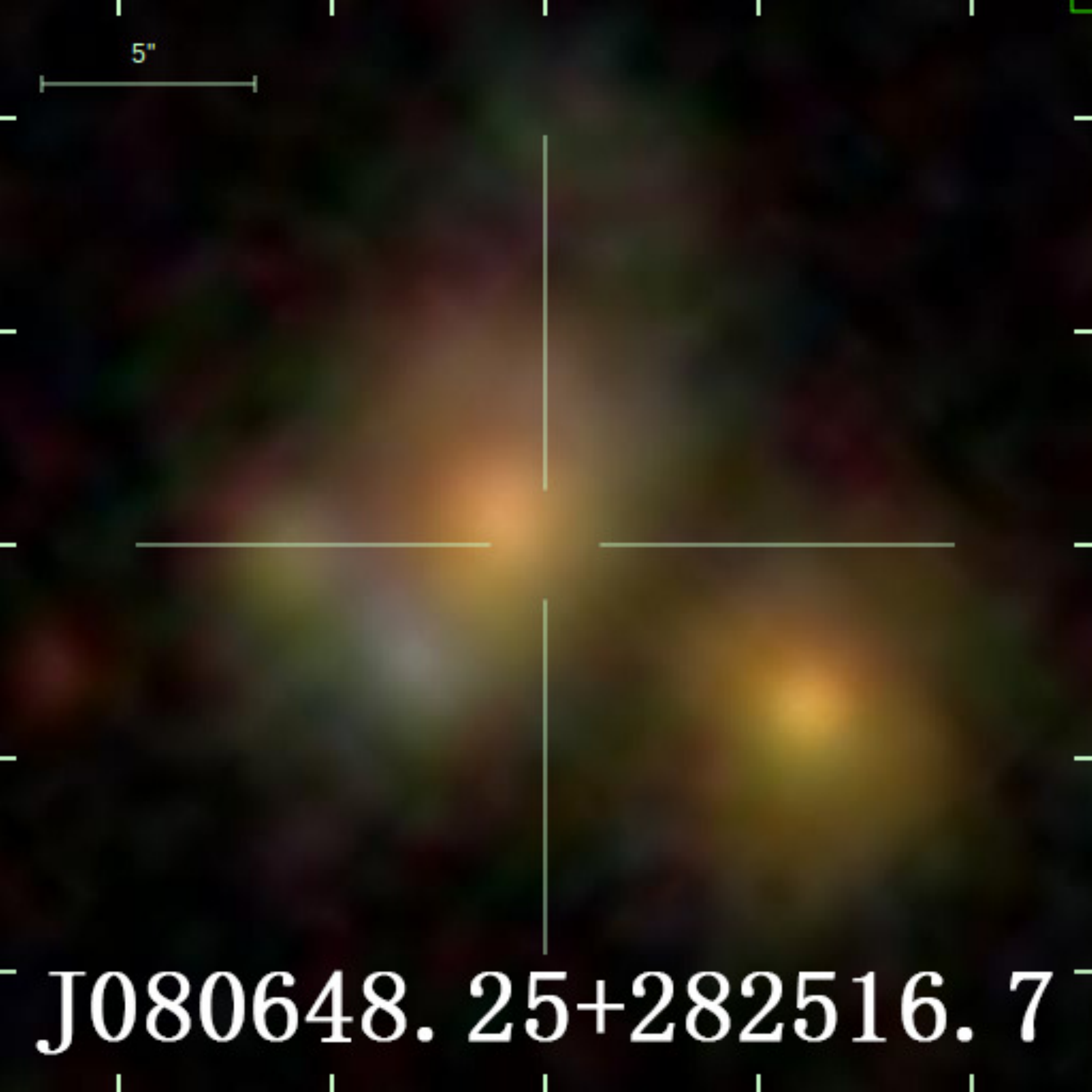}
\includegraphics[width=0.15\textwidth]{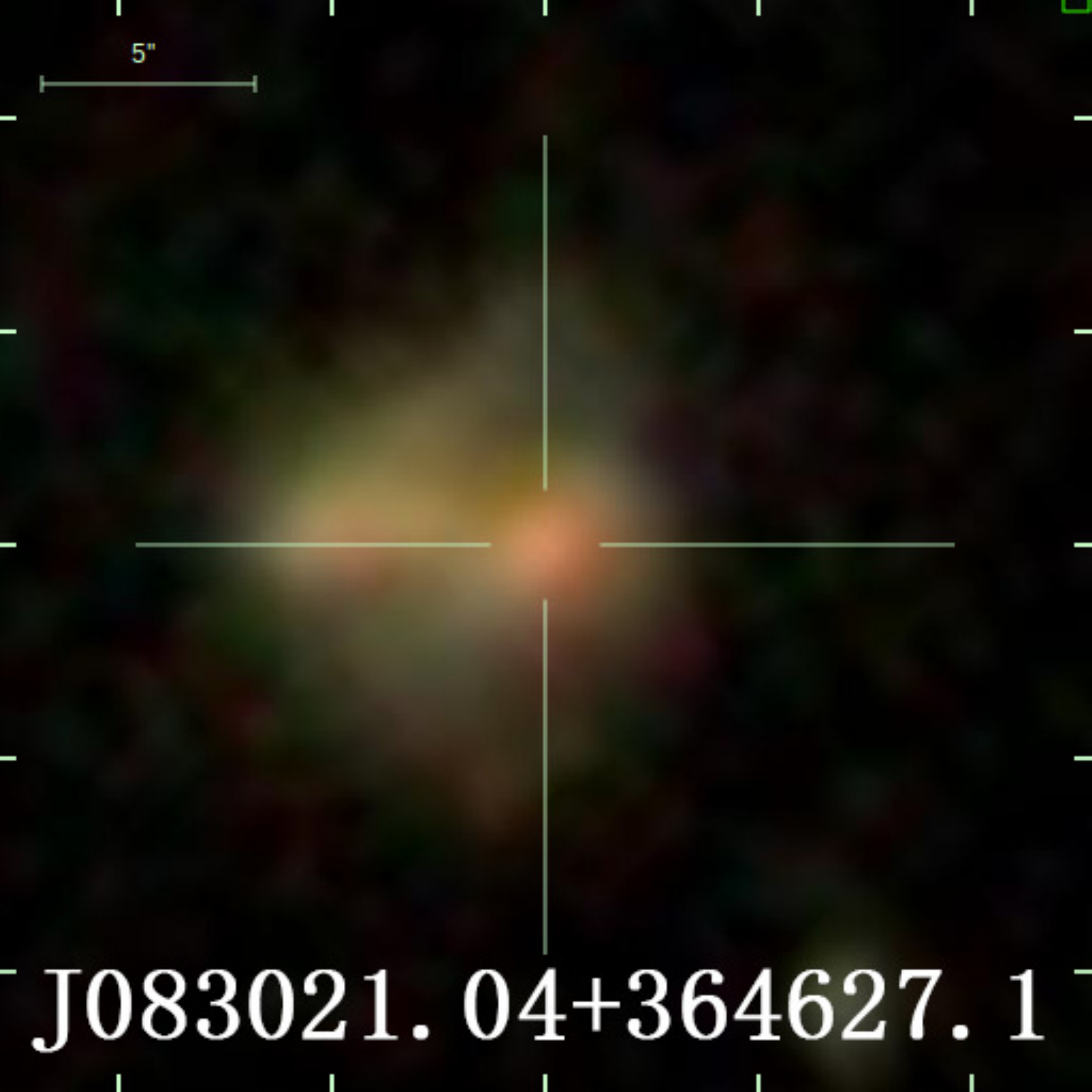}
\includegraphics[width=0.15\textwidth]{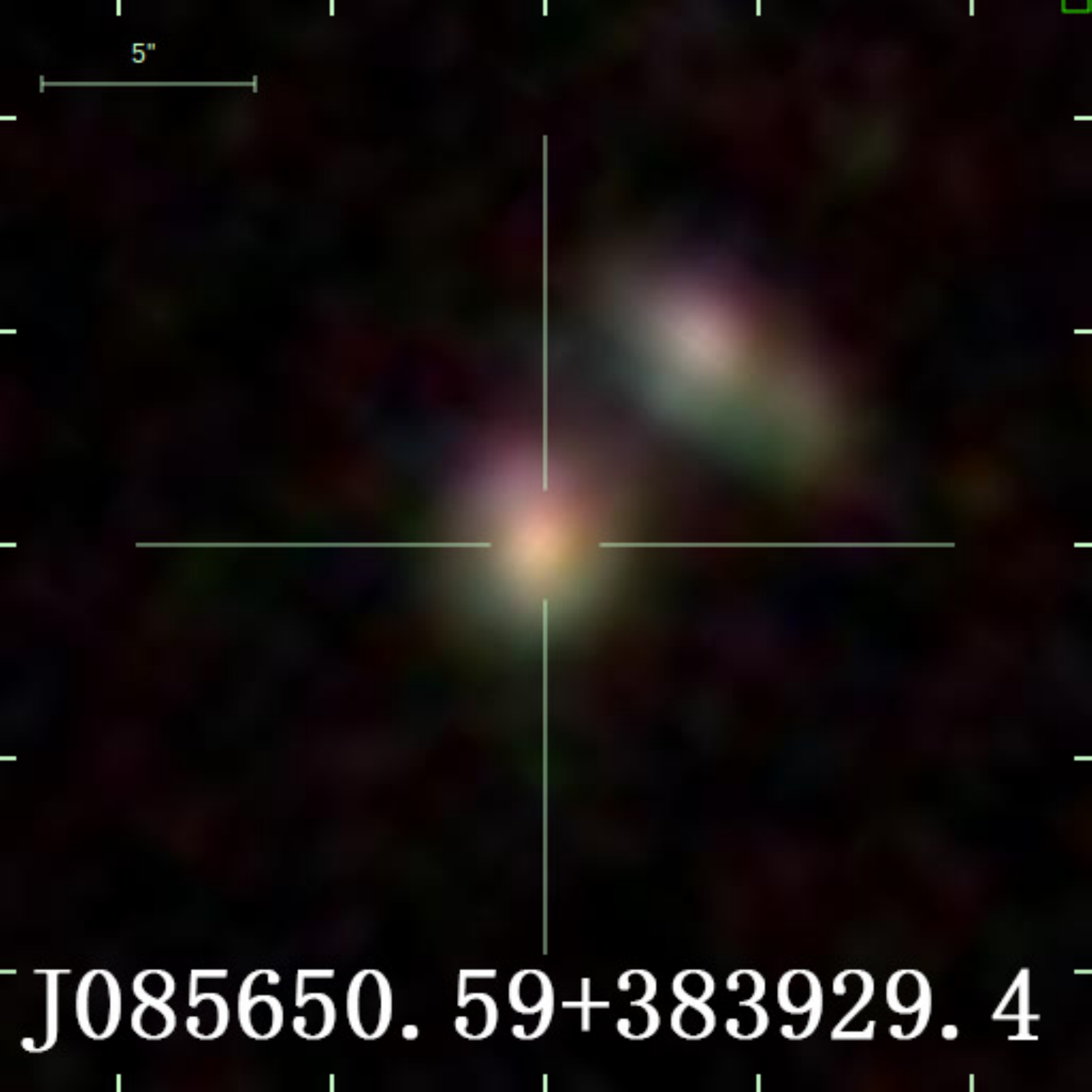}
\includegraphics[width=0.15\textwidth]{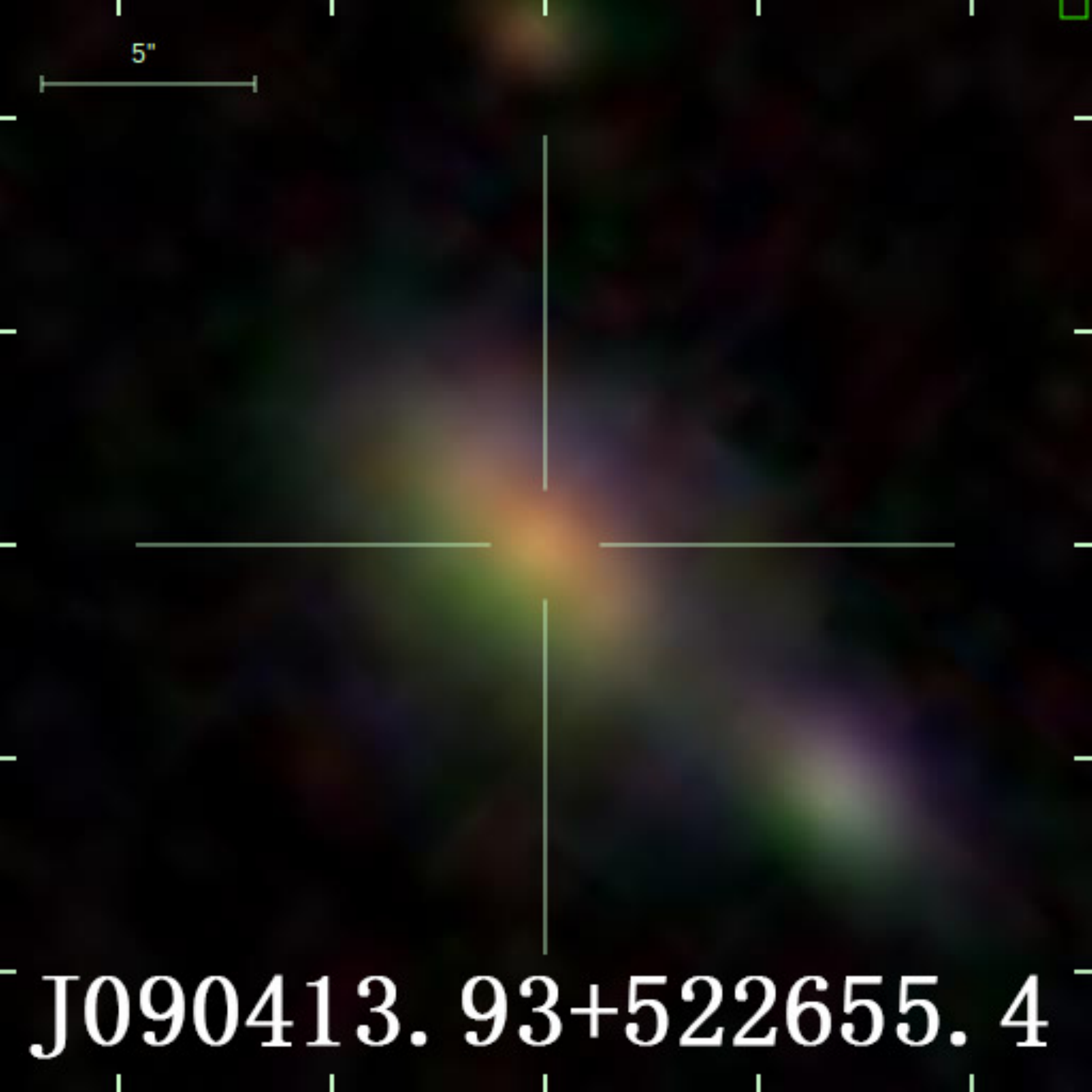}
\includegraphics[width=0.15\textwidth]{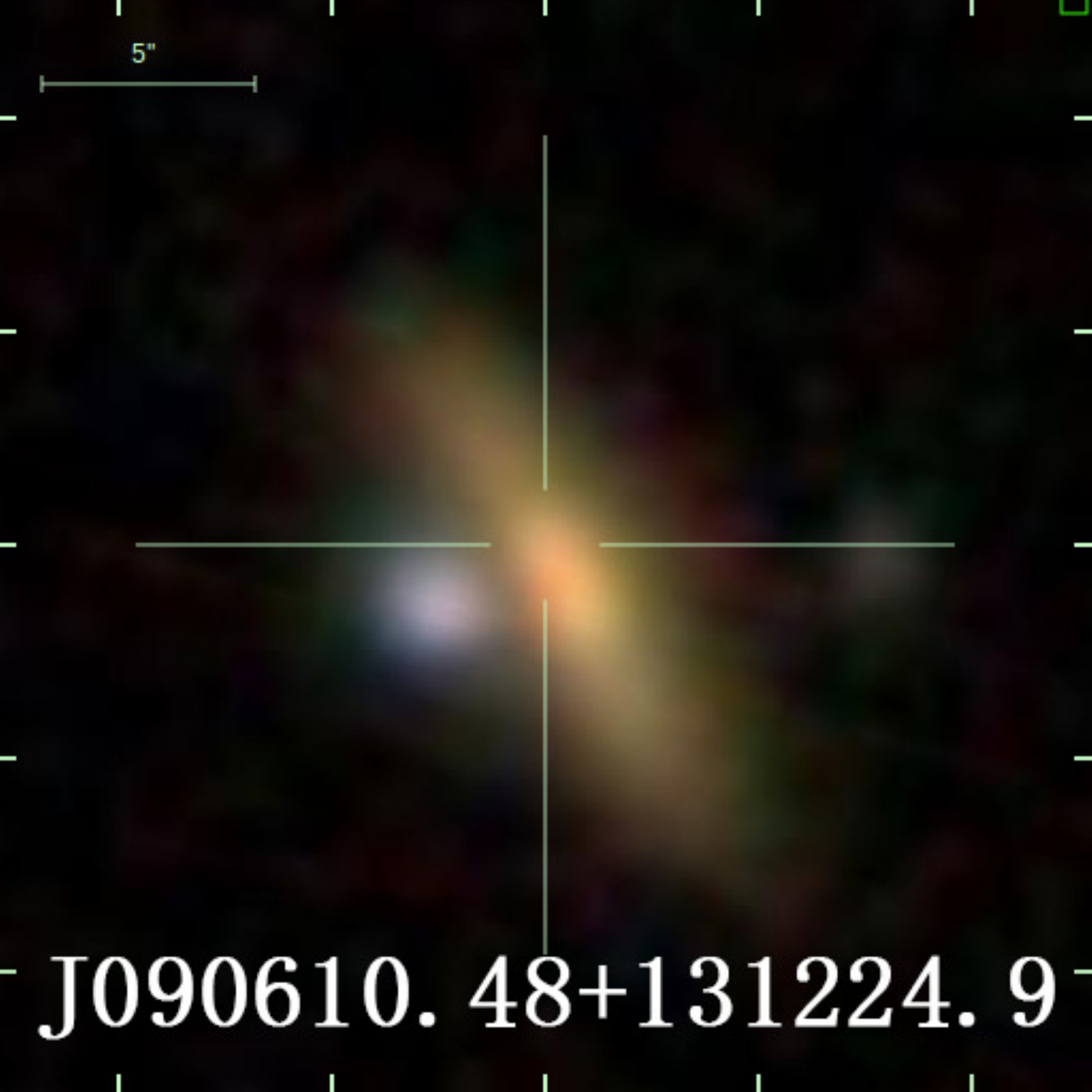}
\includegraphics[width=0.15\textwidth]{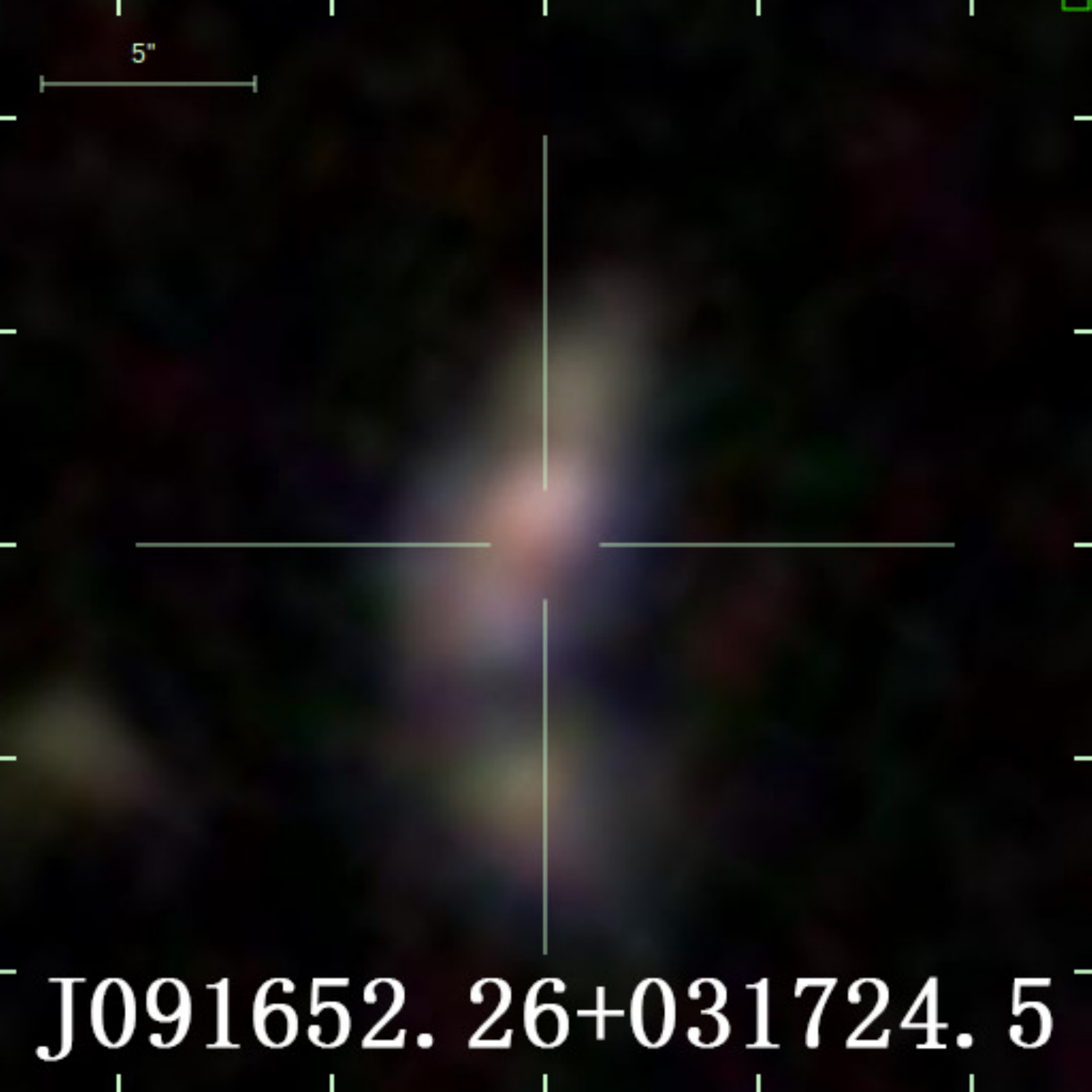}
\includegraphics[width=0.15\textwidth]{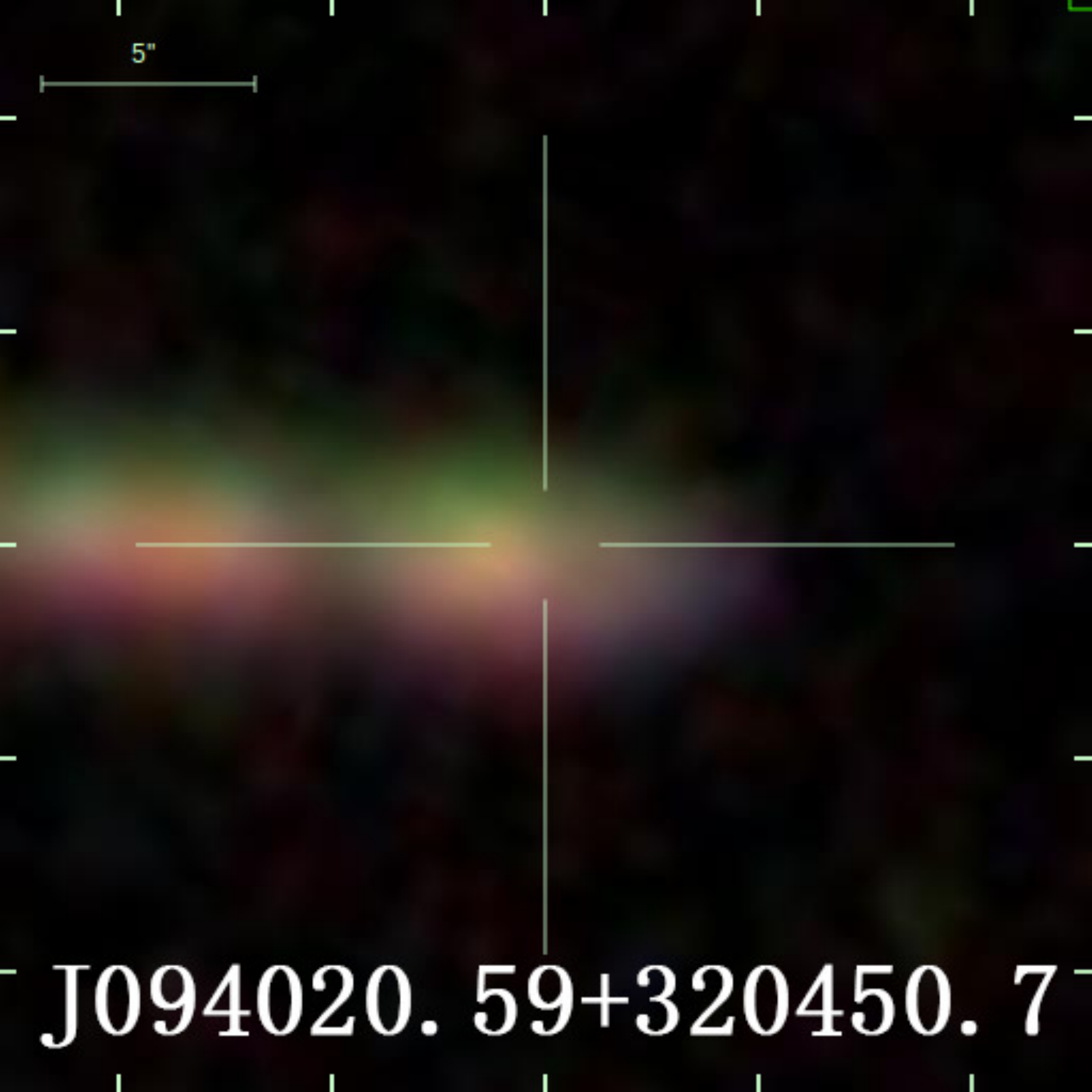}
\includegraphics[width=0.15\textwidth]{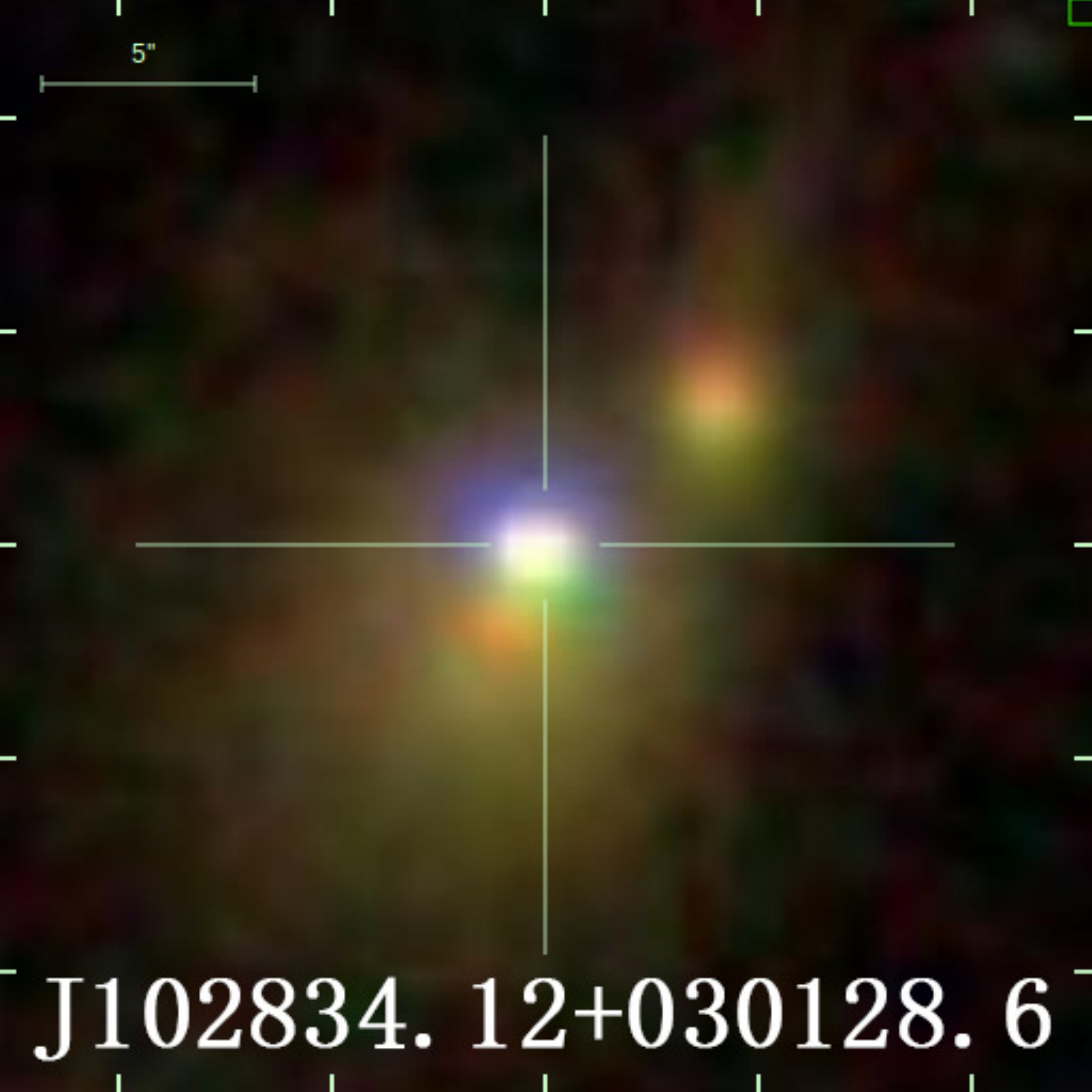}
\includegraphics[width=0.15\textwidth]{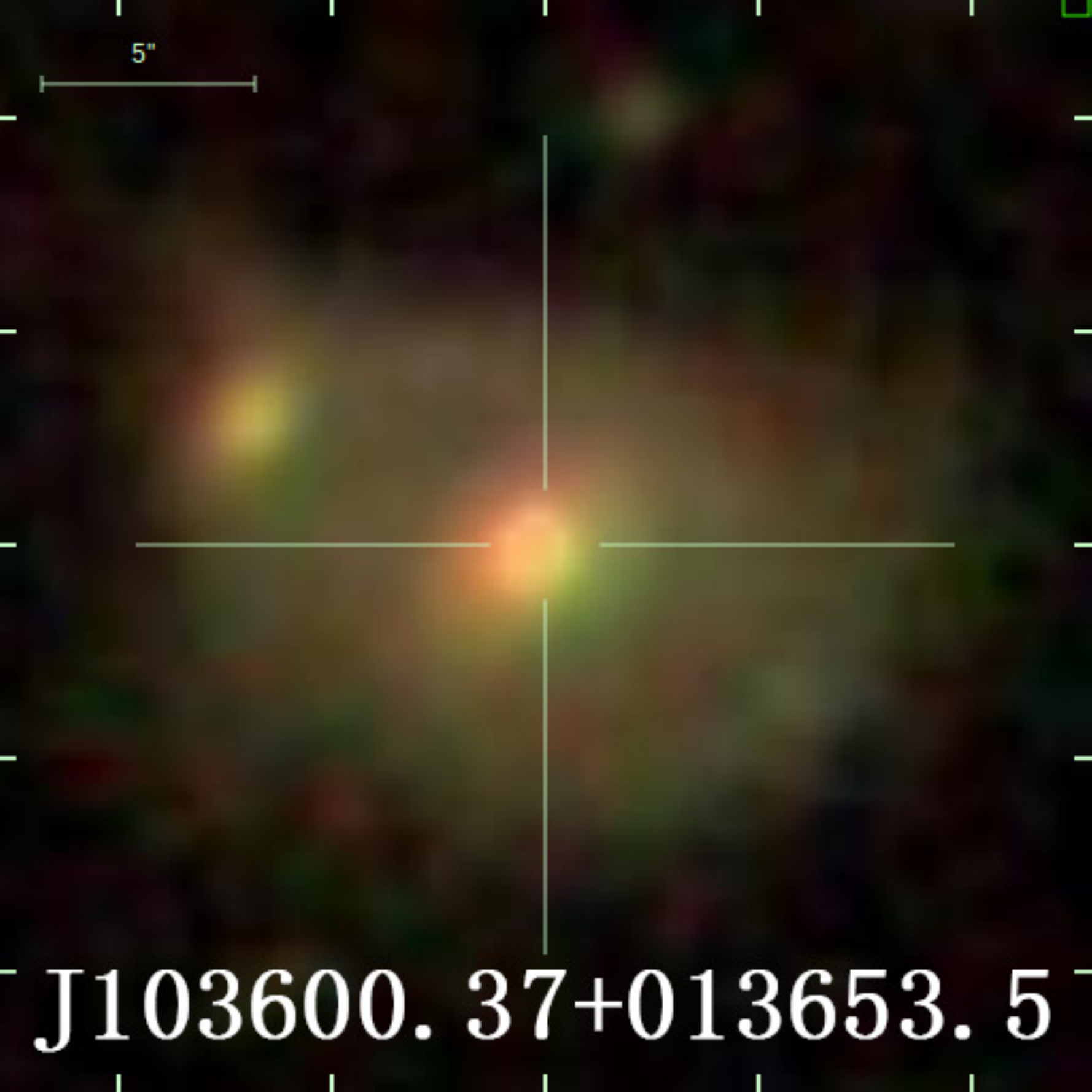}
\includegraphics[width=0.15\textwidth]{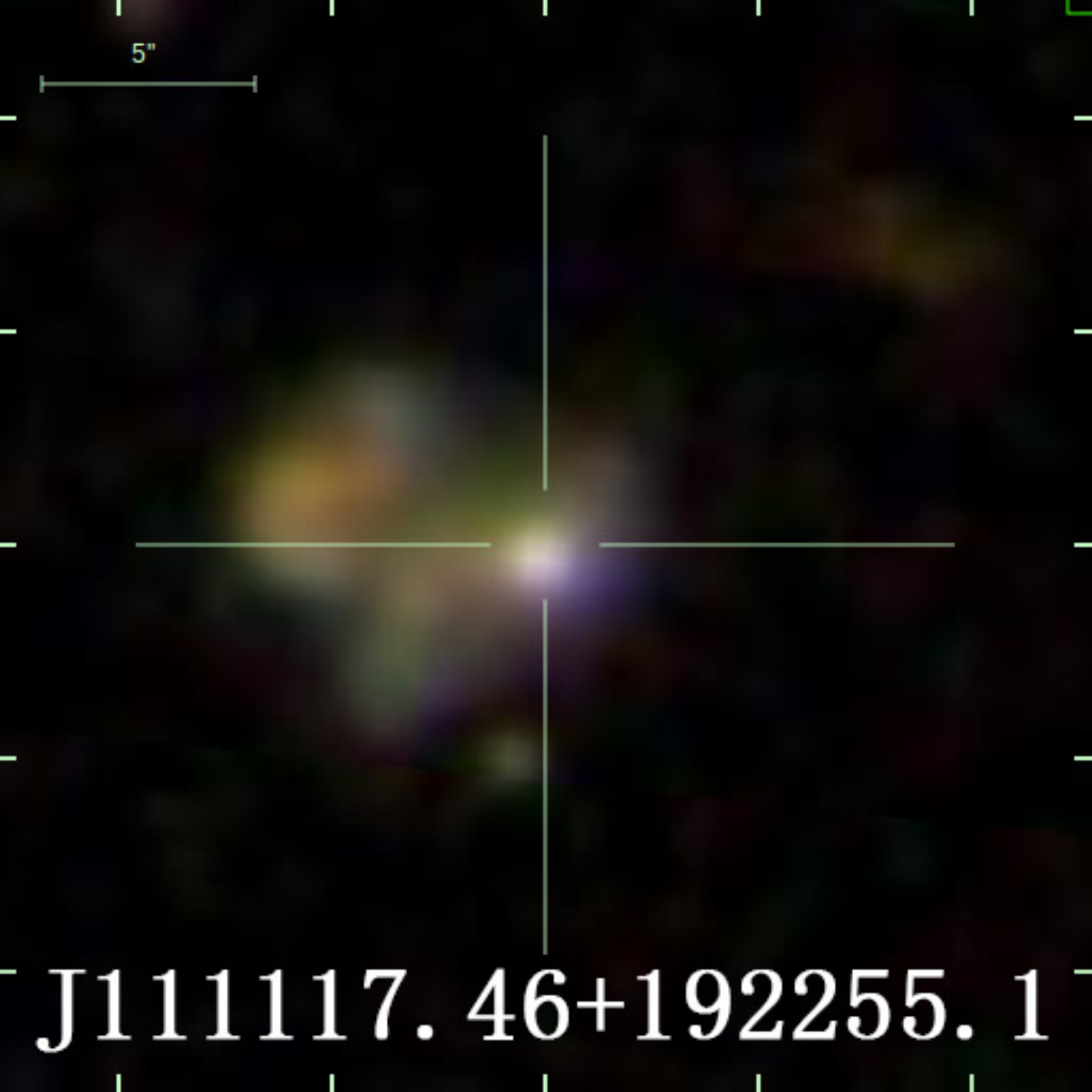}
\includegraphics[width=0.15\textwidth]{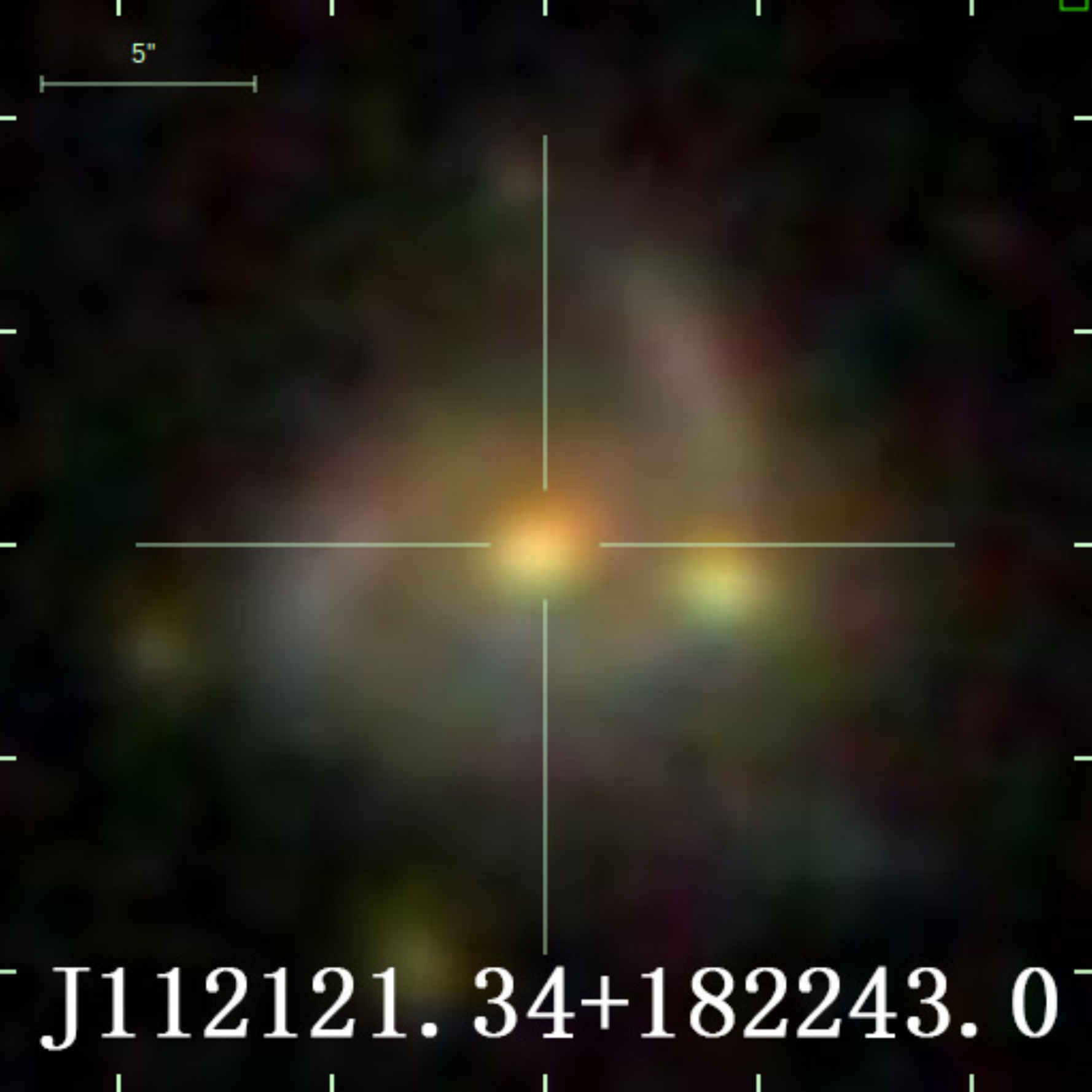}
\includegraphics[width=0.15\textwidth]{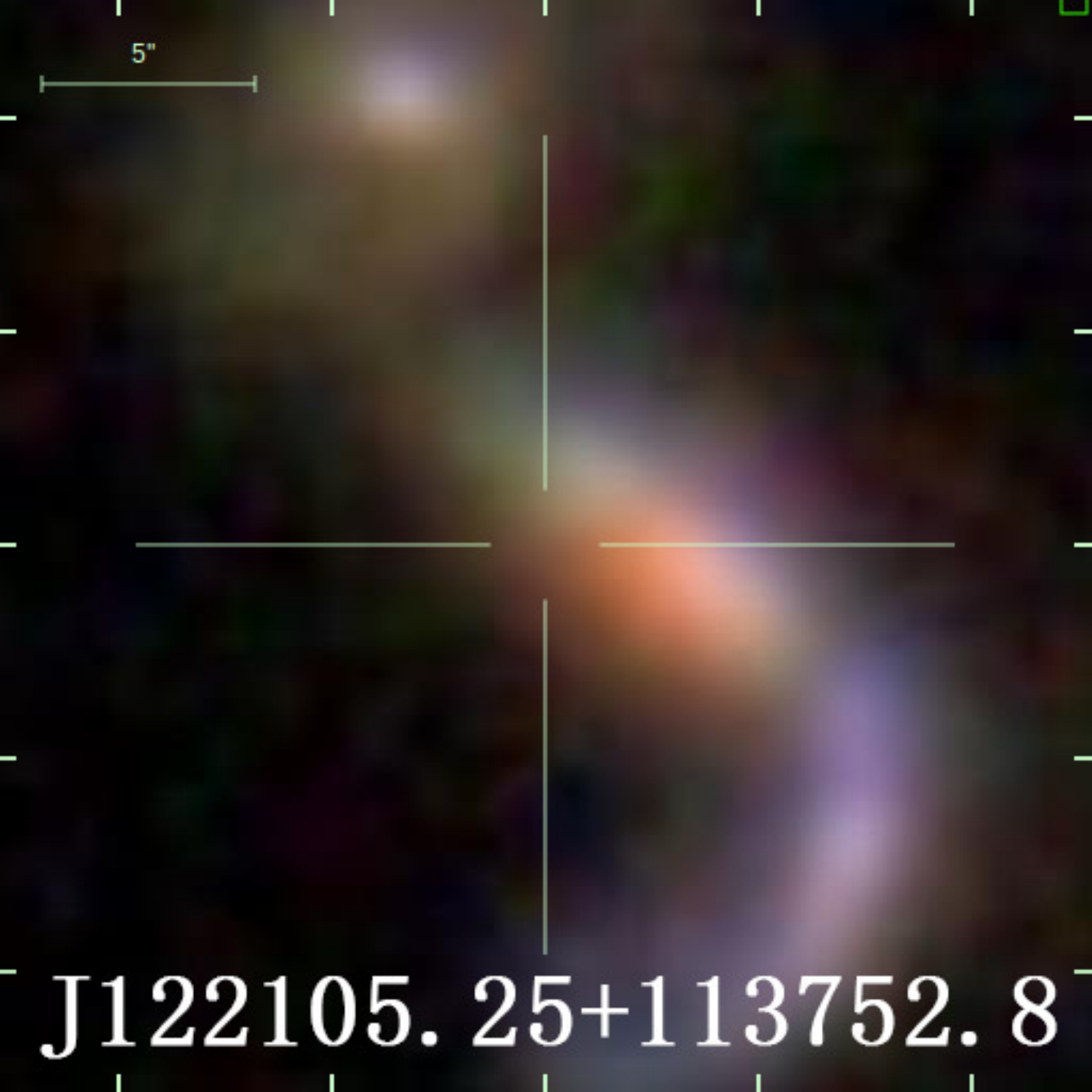}
\includegraphics[width=0.15\textwidth]{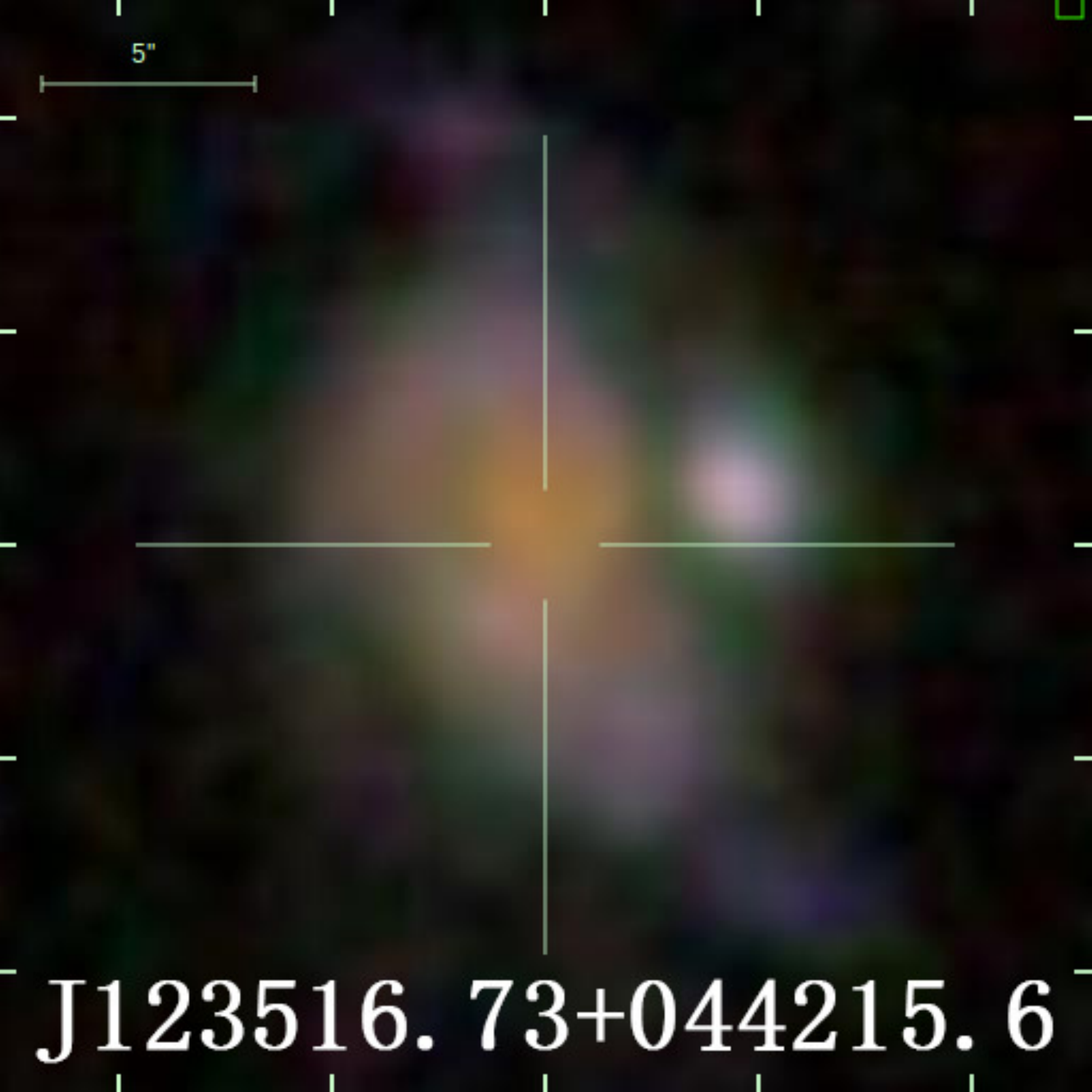}
\includegraphics[width=0.15\textwidth]{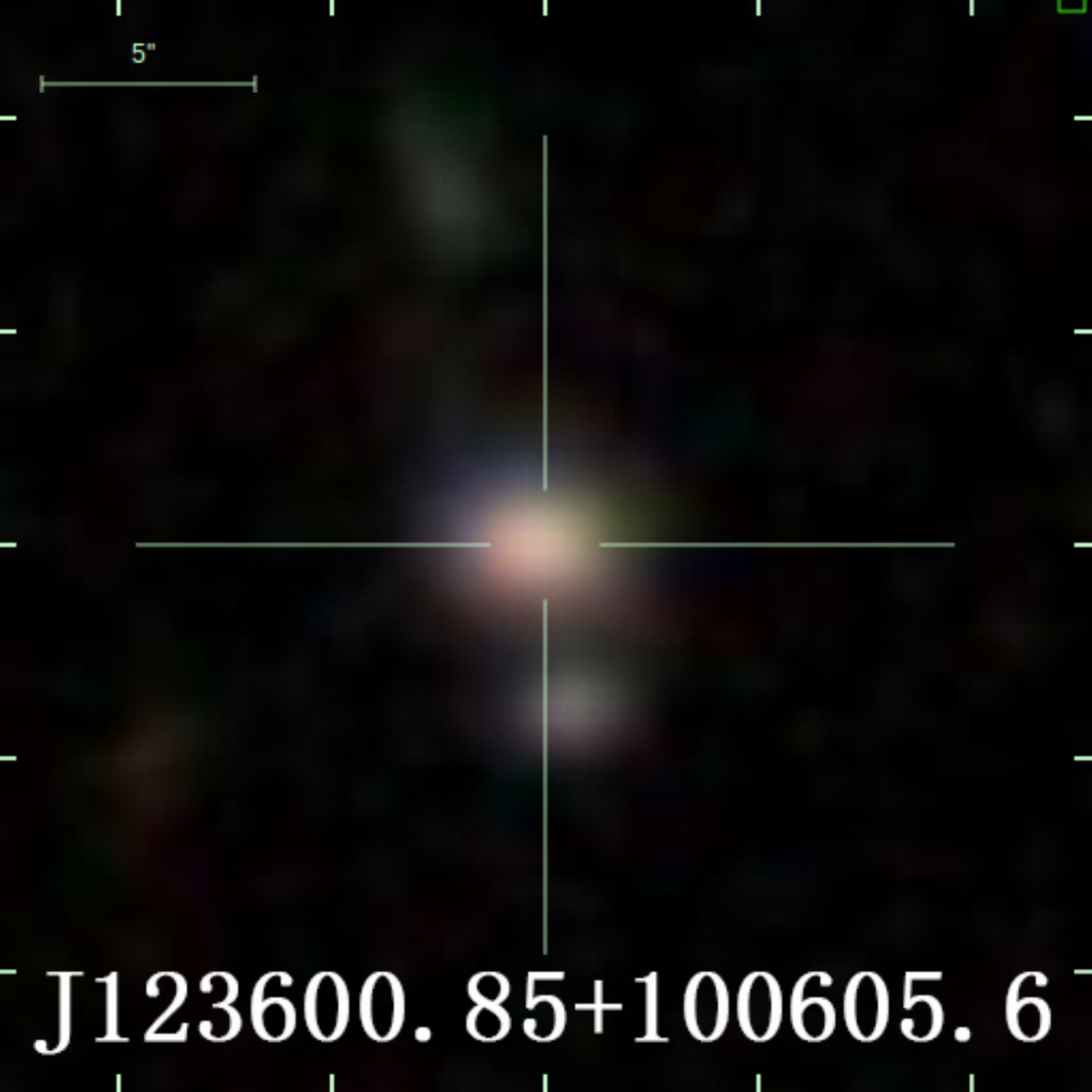}
\includegraphics[width=0.15\textwidth]{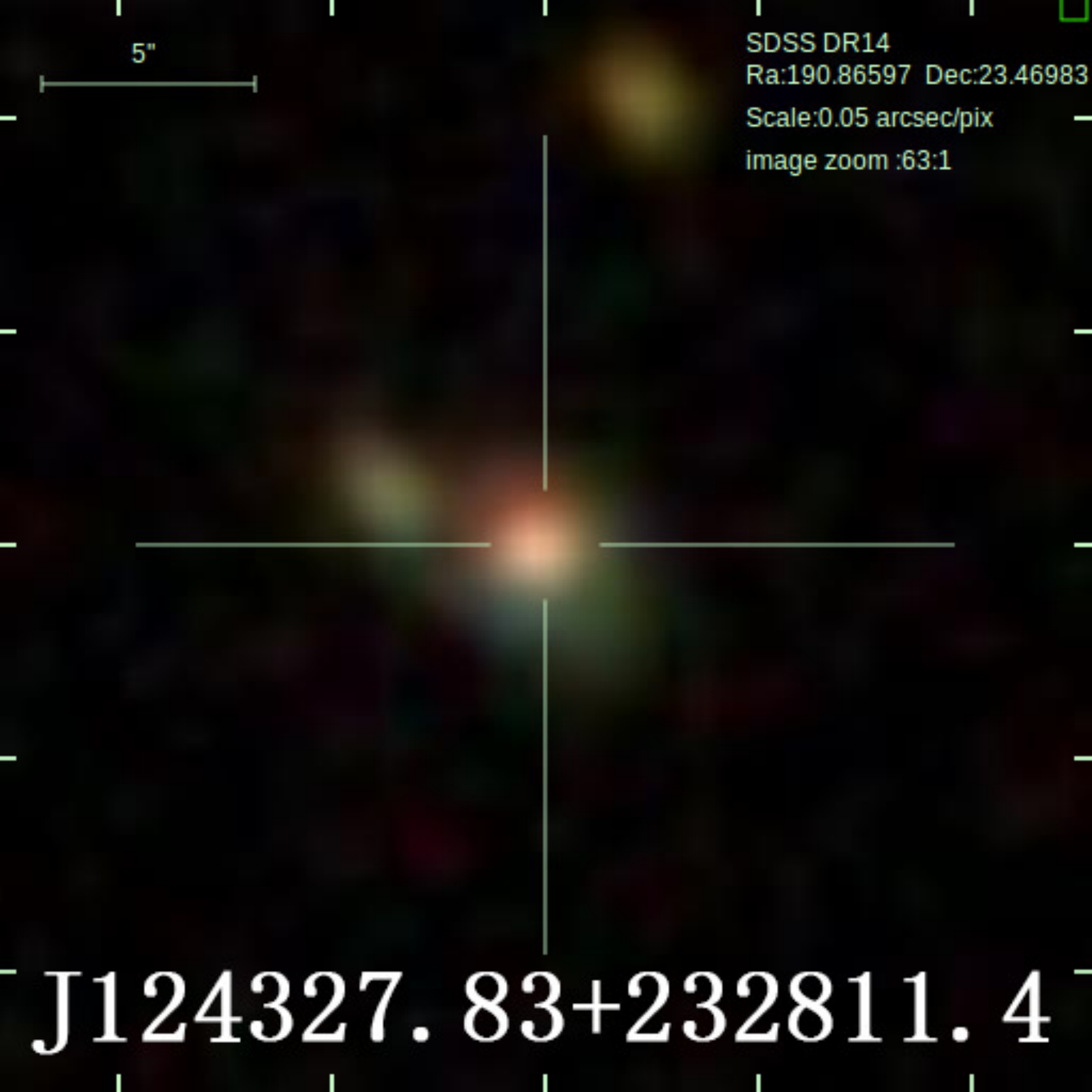}
\includegraphics[width=0.15\textwidth]{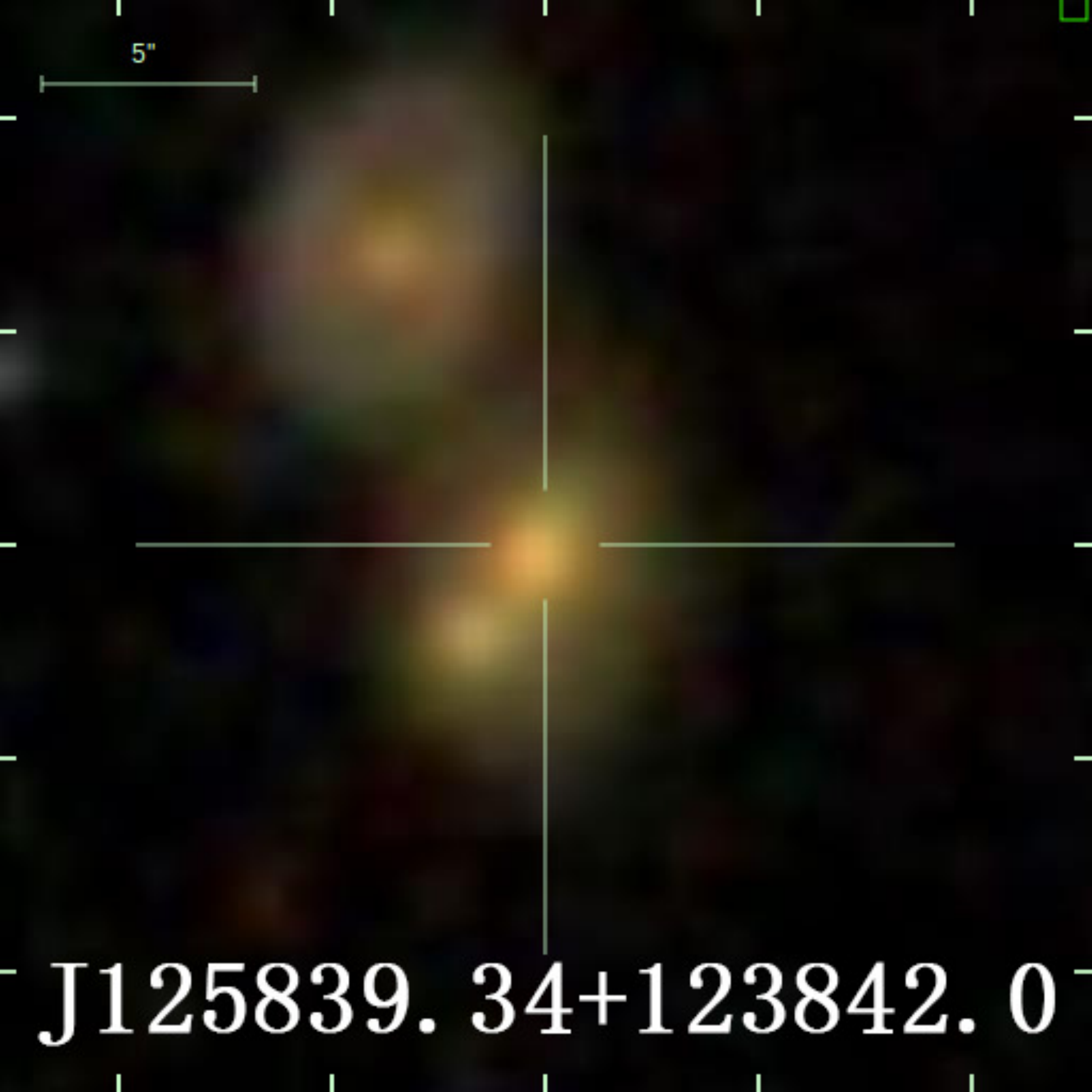}
\includegraphics[width=0.15\textwidth]{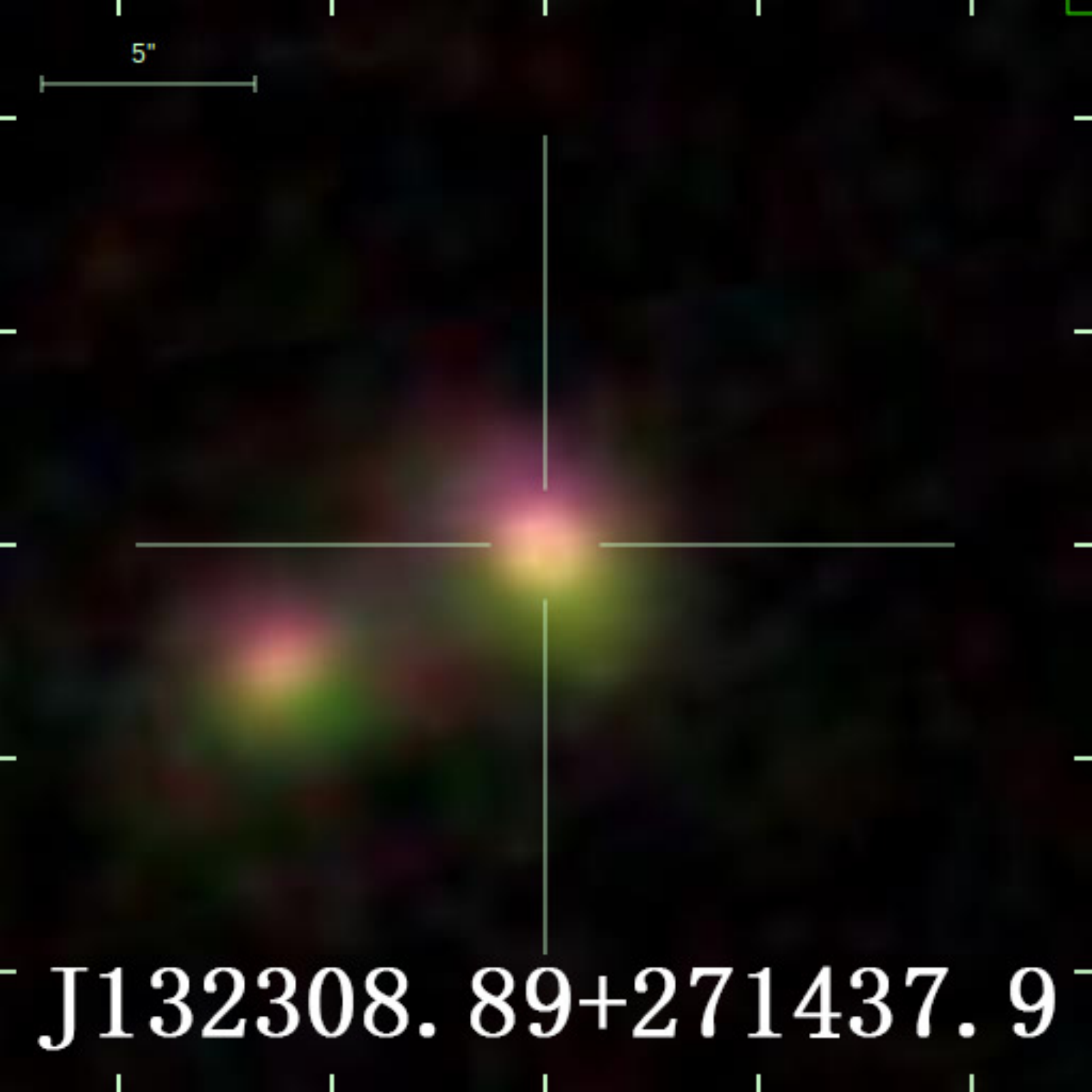}
\includegraphics[width=0.15\textwidth]{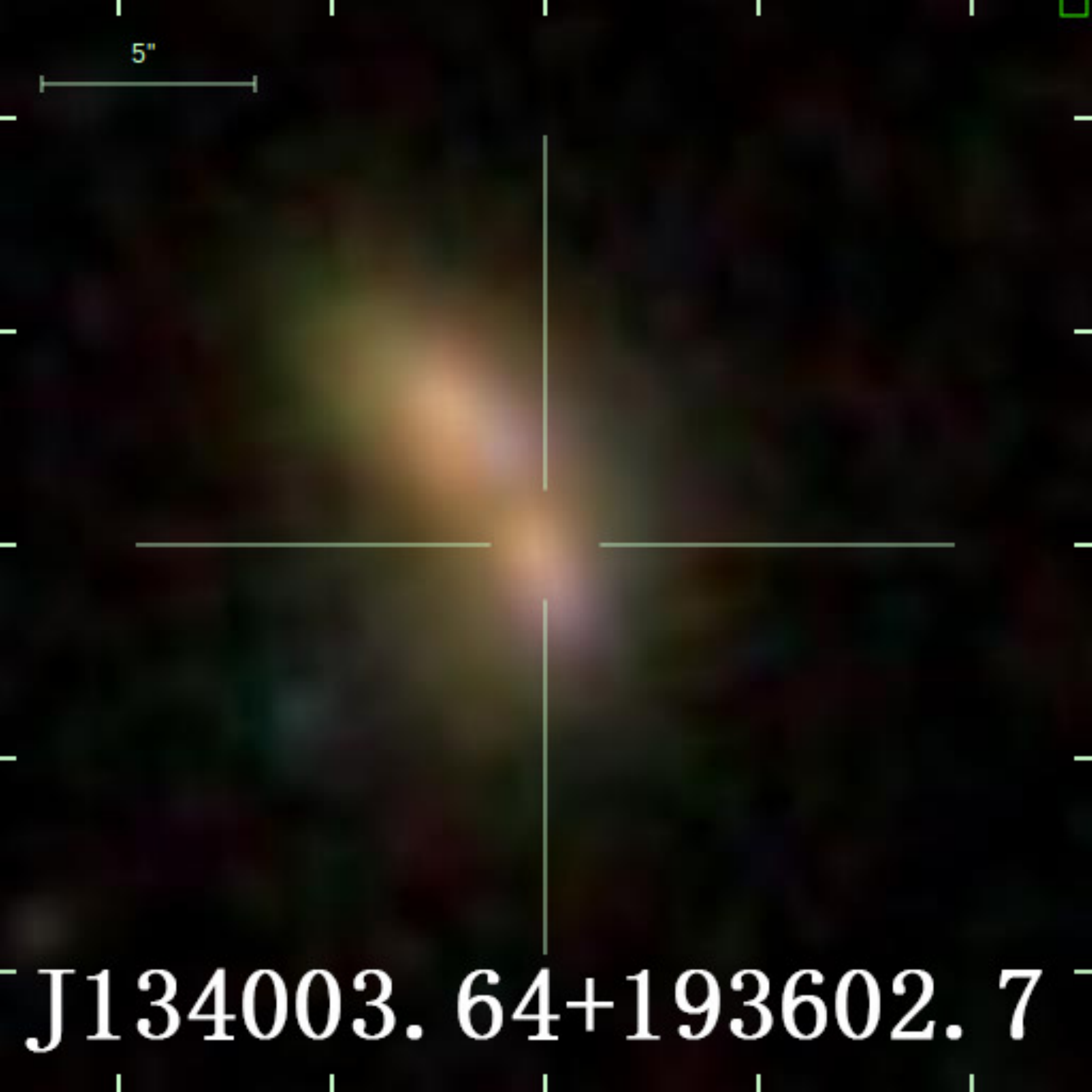}
\includegraphics[width=0.15\textwidth]{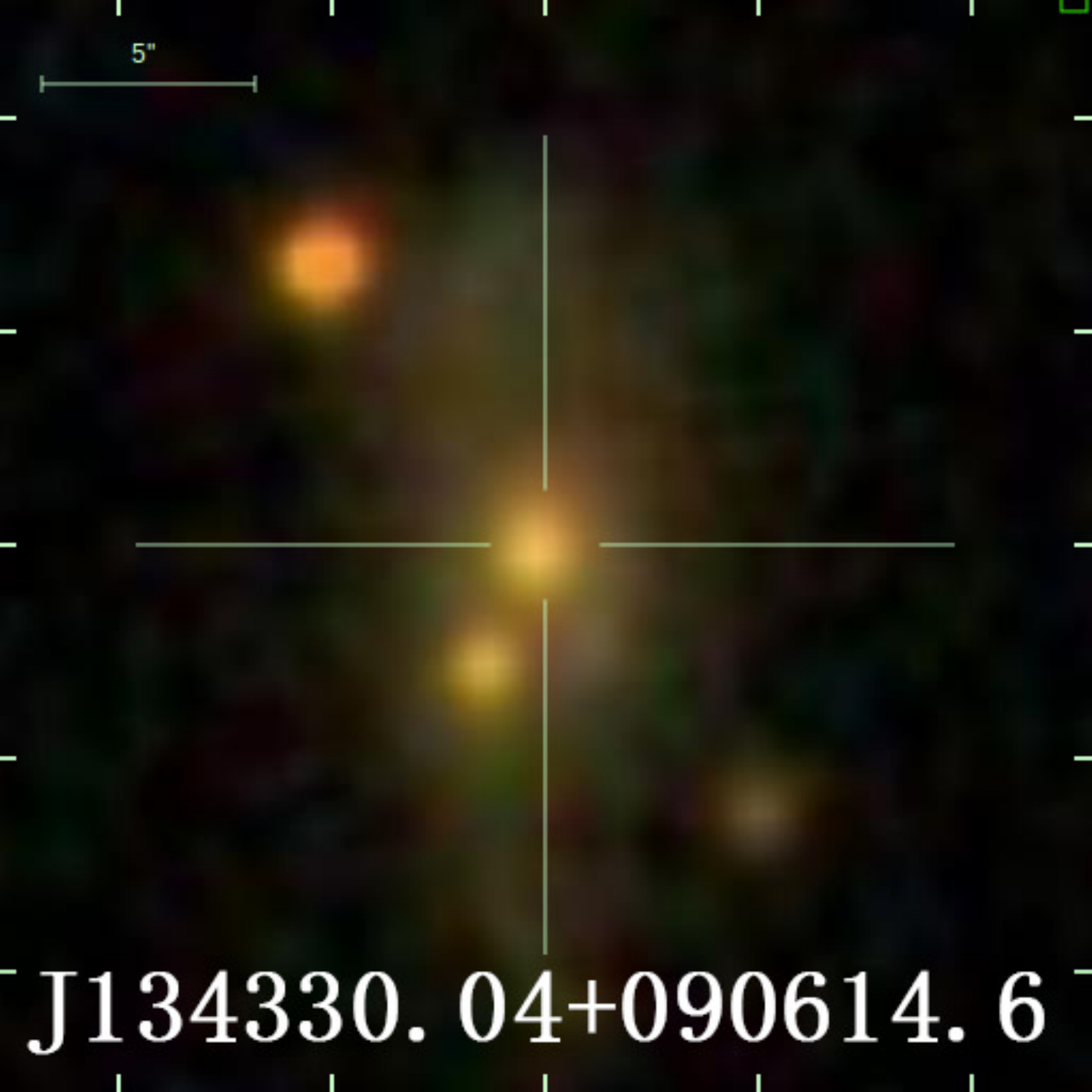}
\includegraphics[width=0.15\textwidth]{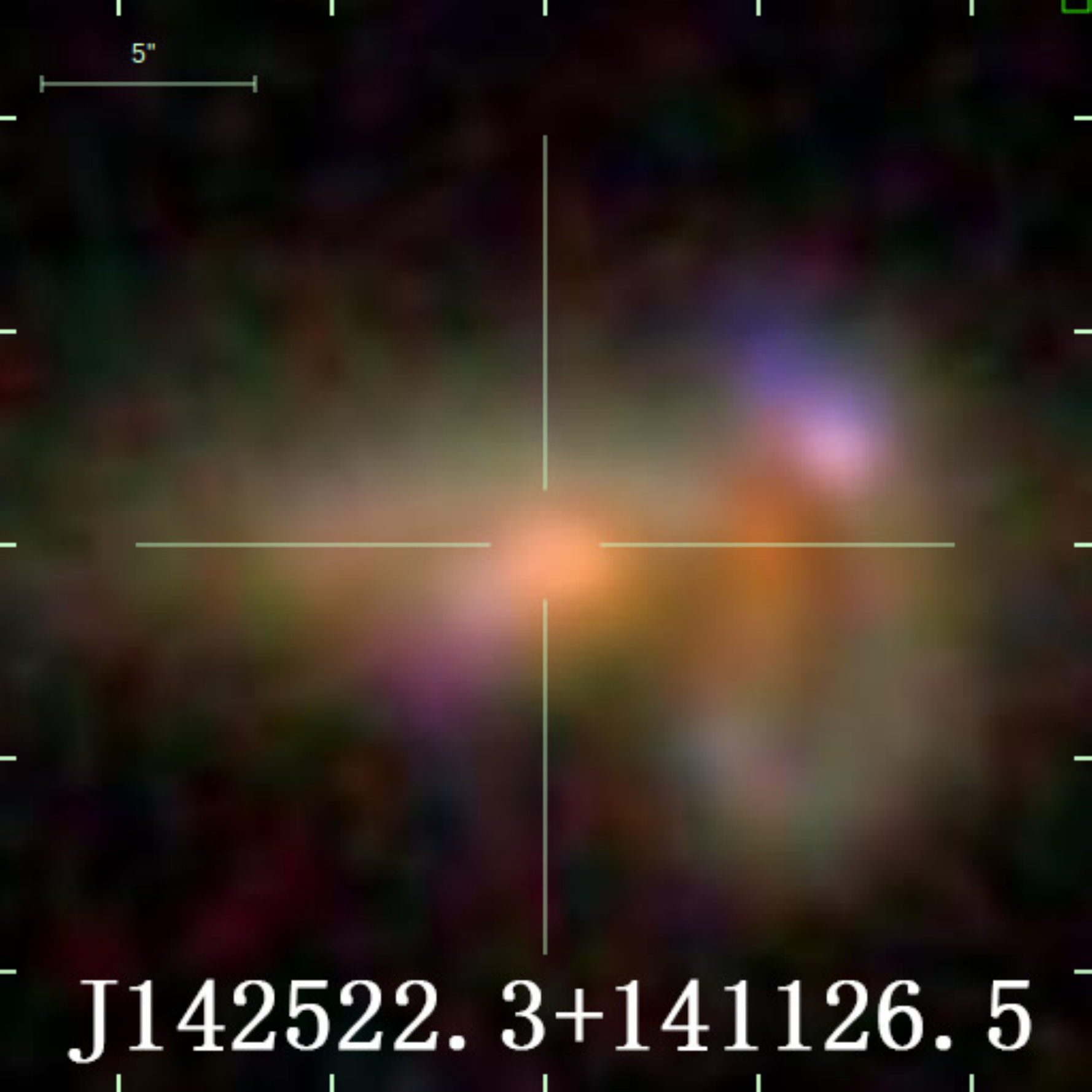}
\includegraphics[width=0.15\textwidth]{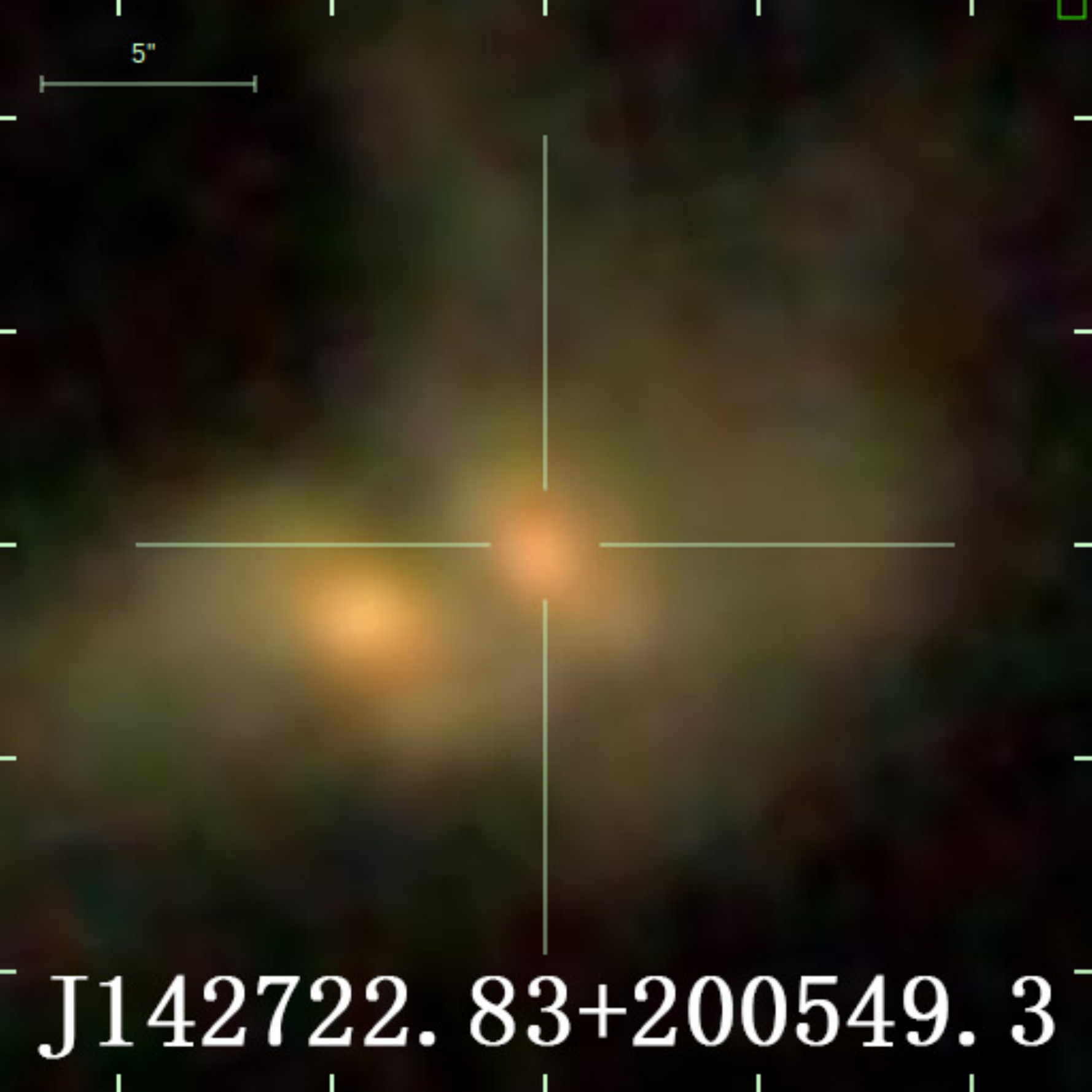}
\includegraphics[width=0.15\textwidth]{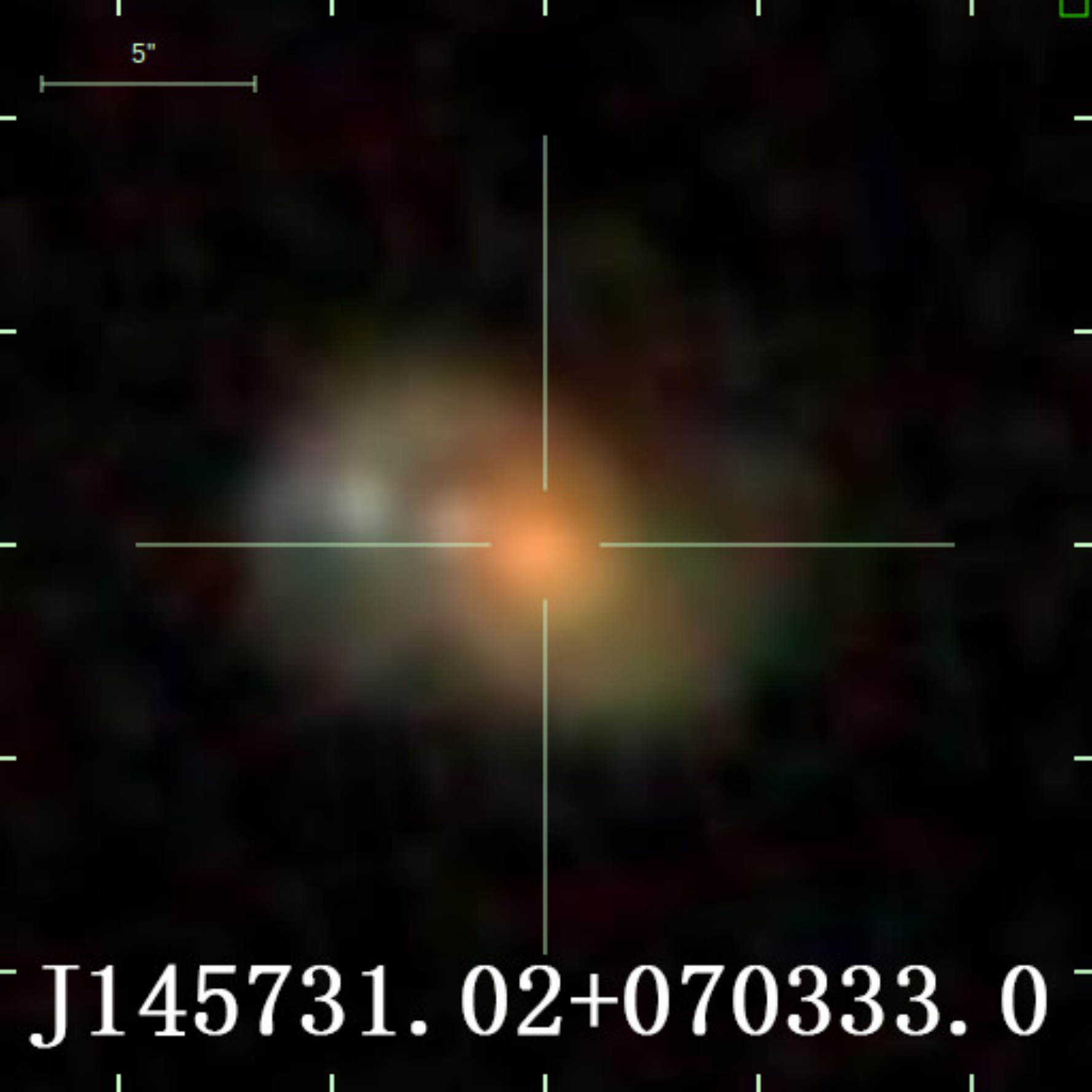}
\includegraphics[width=0.15\textwidth]{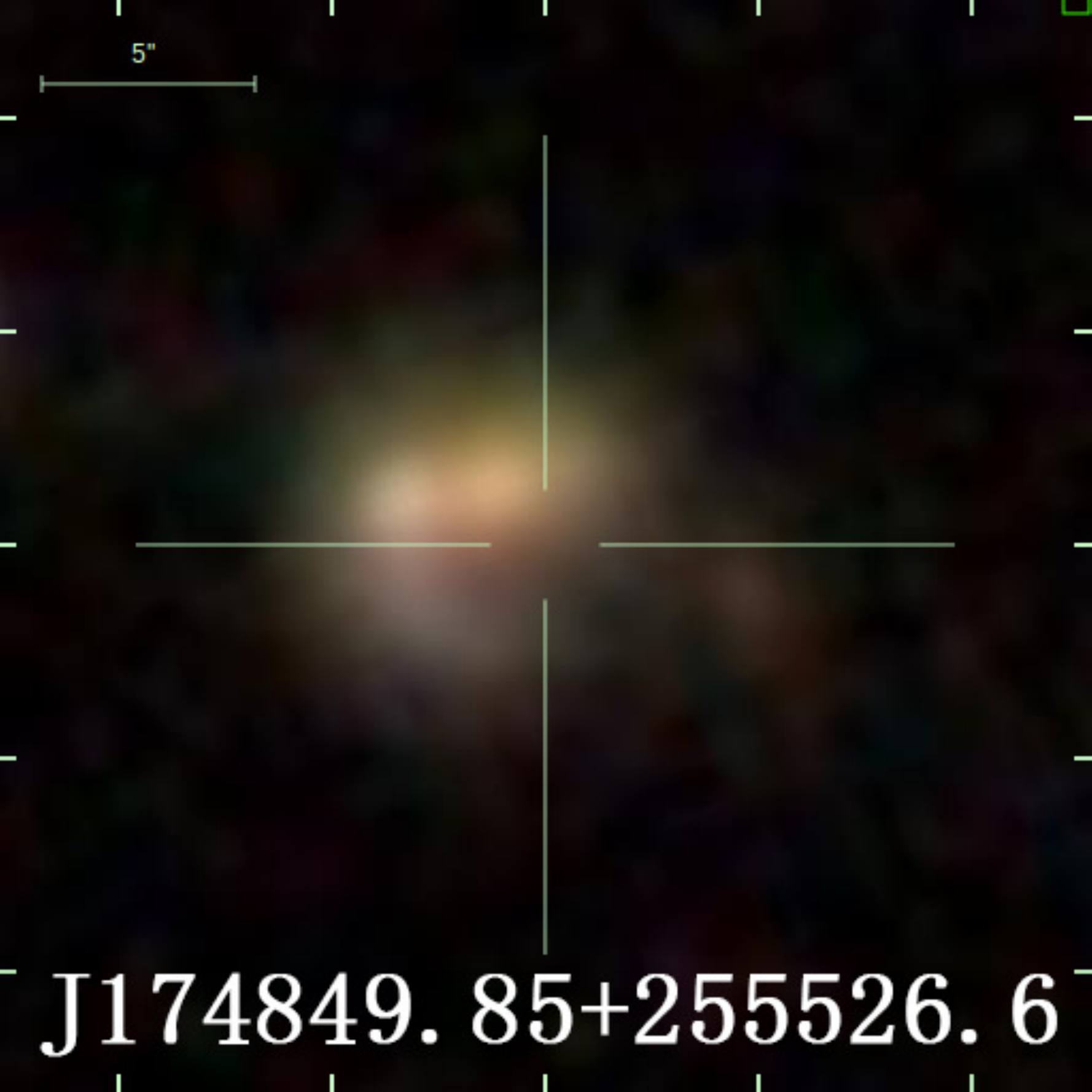}
\includegraphics[width=0.15\textwidth]{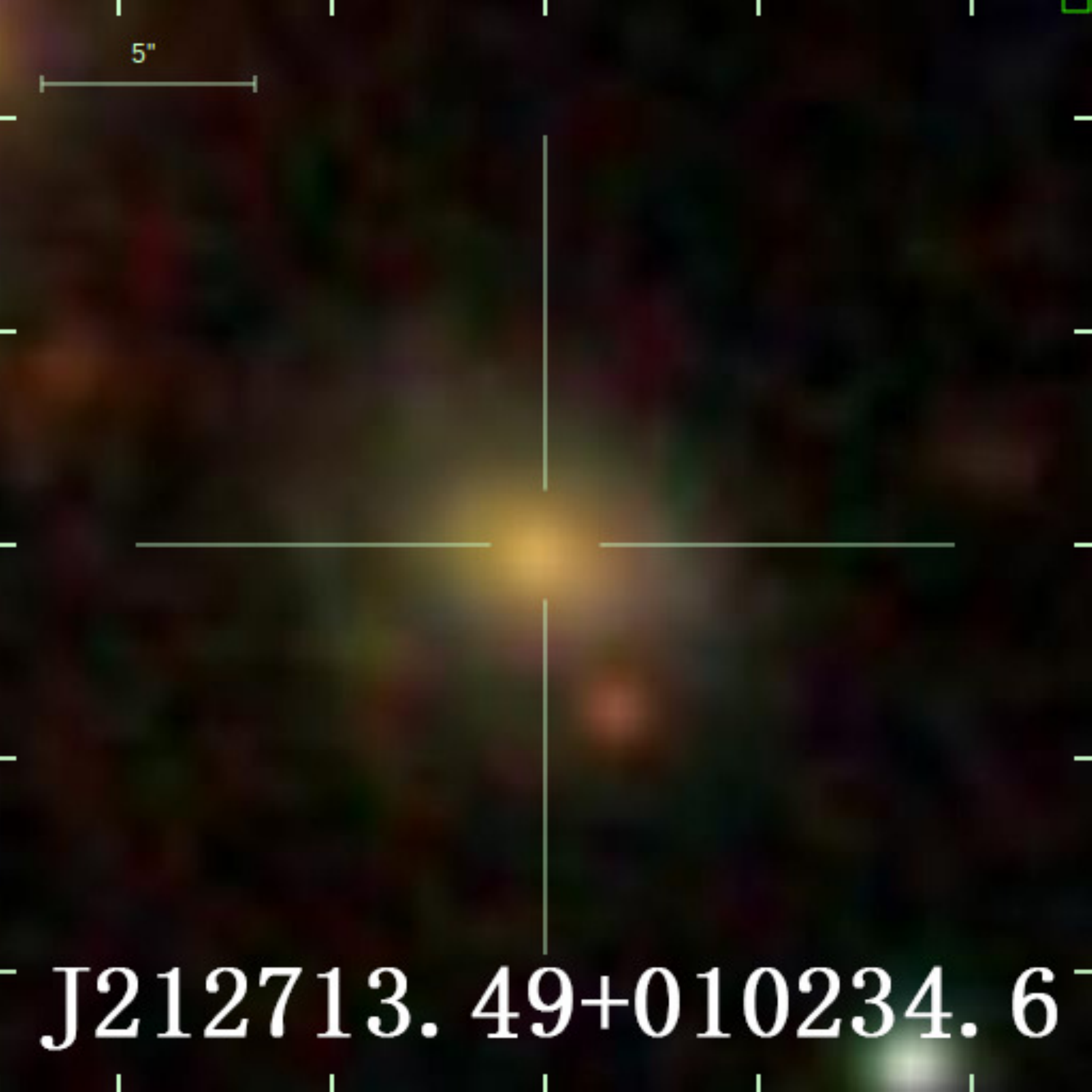}
\includegraphics[width=0.15\textwidth]{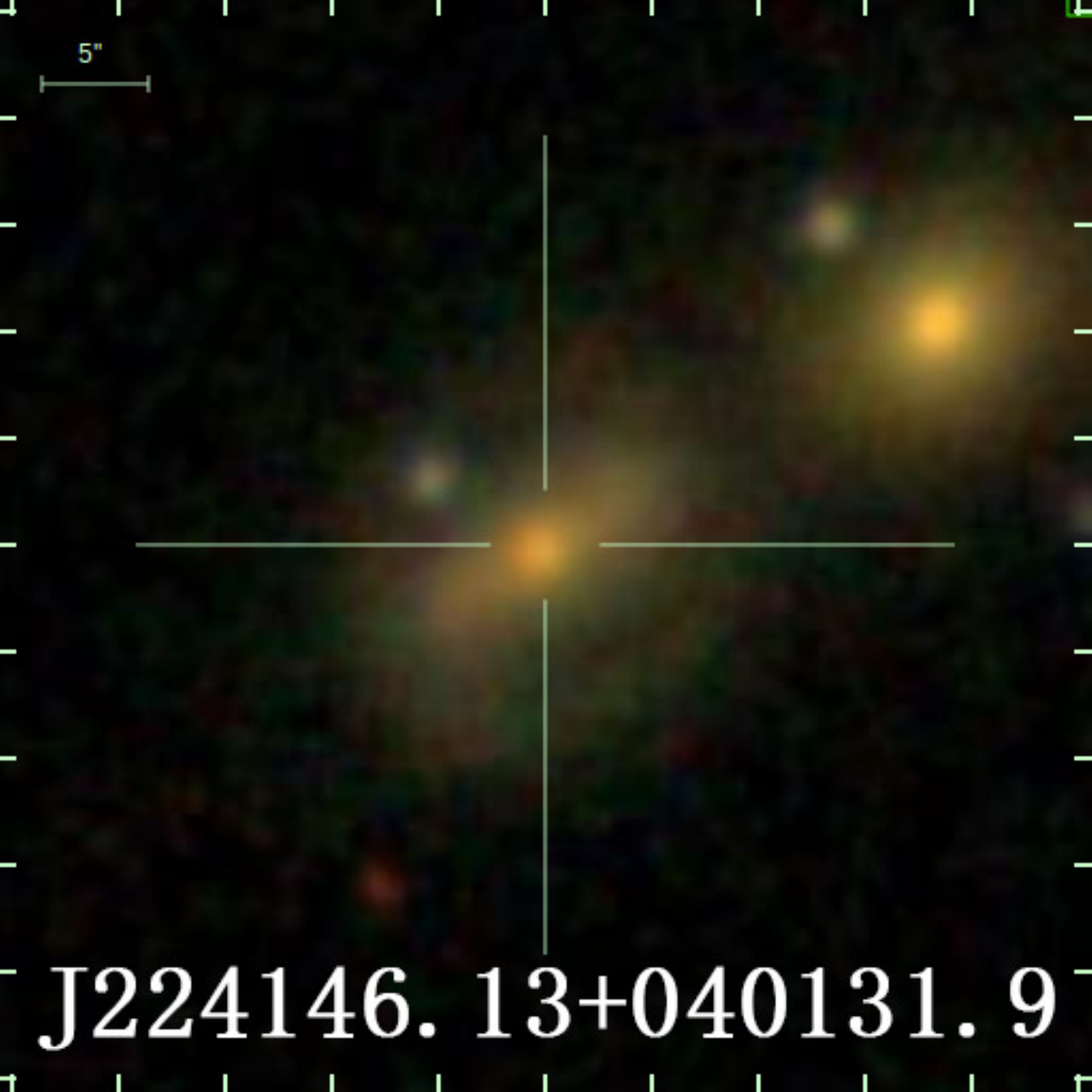}
\caption{SDSS color composite images for 33 candidates, with dual-cored features, close companions or signs of recent interaction. The scale bar of each picture in the upper left corresponds to 5 arcsec. Each picture is centered to the coordinates of LAMOST spectra and has the same orientations (north is up and east is left). LAMOST designation for each target is also tagged.}
\label{SDSS_images}
\end{figure*}

Based on the selected sample, we visually examine all the cross-referenced SDSS images to dig out the galaxies with dual-cored, close companions or other interesting features. Only three of our objects are outside the SDSS photometric footprint. The redshifts of our sample mainly locate around 0.1. At this redshift, projection distance 10 kpc corresponds to $\sim$ 6 arcsec. Considering the SDSS image resolution of $\sim$ 1.4 arcsec, we look for the sources which harbour potential dual cores within $\sim$ 7 arcsec to identify interesting objects, and for the objects whose secondary core has been spectroscopically observed by SDSS, we check the recessional velocity difference between the primary and the secondary core to ensure it is smaller than 500 km s$^{-1}$\citep[e.g.][] {2000ApJ...536..153P}, thus excluding the possible pseudo dual cores caused by projection effects, such as the foreground and background galaxies. The suspected sources with unreliable photometric flags assigned by SDSS are also ruled out. Eventually we find 33 visually interesting sources. Figure \ref{SDSS_images} lists the SDSS multi-color images of these objects, and table \ref{SDSS33_table} displays the related information and analysis notes for these sources. Among these 33 objects, eight targets are spectroscopically observed for the first time by LAMOST, five targets are endowed with both spectra for the primary and secondary cores, three objects have the spectra for the counterpart core with respect to the LAMOST targets, while eighteen sources own the spectroscopic observations both from LAMOST and SDSS surveys. The other unsolved objects in the SDSS images, may also be potential candidates for merging sources at projection distance less than 5 kpc, based on their redshifts. It is worth mentioning that there are 7 dual/offset AGN candidates showing signatures of interactions, which are the most promising dual/offset AGN candidates found in our study, and their designations are emphasized with upper-right symbols in table \ref{SDSS33_table}. These objects amount to a fraction of $\sim$ 9\% with respect to our 77 dual/offset AGN candidates, and considering the non-negligible sample selection bias in two studies, it is basically consistent with the dual AGNs' confirmation rate (3/18) in \citet{2015ApJ...813..103M}, based on optical long-slit spectroscopy and high-resolution VLA multi-band observations. We need stress here that confirmation of dual structures needs unambiguous spatial resolution of the two individual cores, hence, the sample collected here could not fully exclude the potential influence of kinematics of the NLRs and other possible triggering mechanisms. Therefore, high resolution imaging is needed to help confirm the presented observations. 

\section{MaNGA OBSERVATIONS}

\subsection{MaNGA Survey and a Cross-referenced Target}
\label{sec:MaNGA survey} % used for referring to this section from elsewhere

The MaNGA survey is one of the the three projects within SDSS-IV. Unlike previous SDSS surveys which obtained spectra only at the centers of target galaxies, MaNGA enables spectral measurements across the face of each of ~10,000 nearby galaxies thanks to 17 simultaneous ``integral field units" (IFUs), each composed of tightly-packed arrays of optical fibers, it makes spatially resolved maps of individual galaxies using the Sloan spectrographs for the first time. SDSS DR14 has released the second spectroscopic data set comprising  2812 MaNGA galaxies (including ancillary targets and $\sim$ 50 repeated observations) and associated spectrophotometric calibration stars \citep[see][]{2015ApJ...798....7B}.

We cross reference the MaNGA catalog with our selected 325 objects requiring a positional match better than 5 arcsec, and identify a single target, whose unique MaNGA ID is 1-556501. It is selected from the MaNGA primary sample, and is observed with a size-matched IFU in diameter of 12 arcsec (19 fibers) on 2015 November 12. Nine exposures are taken for a total integration time of 8100.87 s. The target data is processed using the standard MaNGA Data Reduction Pipeline (DRP) version 2.1.2. In the LAMOST survey, the designation for this target is J074810.95+281349.2 (OBSID 260303072), and is observed on 2014 November 5.

MaNGA provides spatially resolved spectra, which are complementary to the LAMOST spectra, and together they can enable us to constrain the origin of double peaks for a given galaxy. As has been discussed, the double-peaked narrow emission lines can be produced by kpc-scale dual AGNs, disk rotation and the NLR structure of biconical AGN outflows, while we are unable to directly identify dual AGNs with IFU data alone, we could explore their potential kinematics.

\begin{landscape}
\begin{table}
	%\centering
	\caption{The Sample}
	\label{sample_table}
	\begin{center}
	\begin{tabular}{lcccccccccccccr} % fifteen columns, alignment for each
		\hline\hline
		DESIGNATION & $z$ &$\Delta$V$_{\rm H\beta-[OIII]}$&F$_{\rm H\beta}^{\rm b}$&F$_{\rm H\beta}^{\rm r}$&F$_{\rm [OIII]}^{\rm b}$&F$_{\rm [OIII]}^{\rm r}$&$\Delta$V$_{\rm H\alpha-[NII]}$&F$_{\rm H\alpha}^{\rm b}$&F$_{\rm H\alpha}^{\rm r}$&F$_{\rm [NII]}^{\rm b}$&F$_{\rm [NII]}^{\rm r}$&F$_{\rm H\alpha}^{\rm broad}$ &BPT type& flag\\\hline
		\multicolumn{15}{c}{2-Seyfert II} \\ \hline
		J111201.79+275053.9	&0.04733&	323&319&298& 257 &358 & 317	&
298&422&324 &389& * & 1	&1\\ 
J093144.80-022855.3&0.14409&419&400&327&325&342&390&327&330&345&347&* &1&2\\ 
J103600.37+013653.5&0.10694 &227&320& 263&321& 207 &217& 263 &322&294&295& *&1& 0\\ \hline
		\multicolumn{15}{c}{2-SF} \\ \hline
		J081605.05+185141.2& 0.08081&212&241&227&207&195&199&227&208&220&225&*& 2& 1\\
J013614.92+031356.9& 0.08750& 193&187 & 216 & 223 &205 &204& 216& 216&230&228&*&2& 2\\ 
J141712.05+274315.5& 0.12883& 179&187&198&120& 189& 173 &198& 201 & 200&178 &*&2&0\\ 
J080922.38+174556.1 &0.04489& 182&239& 271& 186 &195&172 & 271&216 &294 &223&*&2& 0\\ \hline
        \multicolumn{15}{c}{2-Liner} \\ \hline
J105103.27+005916.0&0.09163&191&143&143&242&212&191&143&202&272&251&*&3&0\\
J224449.41+154249.6&0.08487&279&374&374&361&199&279&374&347&338&412&*&3&2\\
J092333.10+131059.1&0.09027&238&245&245&195&167&238&245&315&308&386&*&3&1\\ \hline     
        \multicolumn{15}{c}{Broad-line AGN} \\ \hline
 J091544.18+300922.1& 0.12916&278 & 367& 336& 283& 284& 279 &336 & 242 &273& 291 & 1757& 4 & 1 \\
J085821.10+331631.0& 0.07338 &234&322 & 321&215& 215&217&321 & 250& 291 & 268 & 1733 & 4 & 0\\
J164840.15+425547.6& 0.12892& 178 & 157 & 206& 238& 298 &174 & 206 & 280&181 & 310& 2495 & 4&0\\
J141648.46+223520.1& 0.14090&244 &265 & 425& 180&357 &237& 425&362&130&257&2689& 4 & 0\\ \hline
        \multicolumn{15}{c}{SF+Seyfert II} \\ \hline
J013219.98-014505.7&	0.07748&	177&	124&	237&	238&	212&	183&	237&	277&	239&	266&	*&	5&	0\\
J090021.12+524620.4&	0.08524&	234&	211&	268&	262&	148&	208&	268&	204&	285&	219&	*&	5&	1\\
J084204.59+420251.1&	0.13031&	186&	261&	214&	283&	171&	206&	214&	208&	148&	224&	*&	5&	0\\ \hline\hline
	\end{tabular}\\
	\end{center}

Notes. $^{\rm b}$ and $^{\rm r}$ denote blueshifted and redshifted components. Column 1: LAMOST DR4 designation hhmmss.ss+ddmmss.s (J2000.0). Column 2: Redshift of the spectral observation. Column 3: The velocity offset between the blueshifted and redshifted components of H$\beta$-[O~{\sc iii}] lines set relative to the systemic redshift, in units of km s$^{-1}$. Column 4-7: FWHMs of H$\beta$ and [O~{\sc iii}]$\lambda\lambda$4959,5007, respectively, in units of km$^{-1}$. Column 8: The velocity offset between the blueshifted and redshifted components for H$\alpha$-[N~{\sc ii}] lines set relative to the systemic redshift, in units of km s$^{-1}$. Column 9-12: FWHMs for H$\alpha$ and [N~{\sc ii}]$\lambda\lambda$6548,6584, respectively, in units of km s$^{-1}$. Column 13: For broad-line AGNs, the FWHMs of broad H$\alpha$ lines are provided. Column 14: The BPT classification types of the targets: the number from 1 to 8 represents the types 2-Seyfert II, 2-SF, 2-Liner, Broad-line AGN, SF+Seyfert II, SF+Liner, Seyfert II + Liner and unknown separately. Column 15: flag for previously report, 0 represents that the source has already been spectroscopically observed by other surveys but was missed by the previous searches, 1 stands for the one has been reported as dual-peaked galaxy candidate\citep[see][]{2012ApJS..201...31G}, and 2 means that the target is spectroscopically observed for the first time by LAMOST. Since LAMOST records only the relative flux, the flux of the lines is not listed. The average uncertainty for systemic redshift is $\sim$ 17 km s$^{-1}$. For H$\beta$-[O~{\sc iii}] and H$\alpha$-[N~{\sc ii}] lines set, the average statistical errors in the best-fit velocity offsets are $\sim$ 21 and $\sim$ 20, respectively, in units of km s$^{-1}$, the average statistical errors in the best-fit FWHMs are $\sim$ 11,  $\sim$ 15,  $\sim$ 9, $\sim$ 9 for H$\beta$, [O~{\sc iii}], H$\alpha$ and [N~{\sc ii}] lines, respectively, in units of km s$^{-1}$.
(A portion is presented here for guidance regarding its content and form.)
 \end{table}
 \end{landscape}

\begin{table}
	\caption{The 33 objects of interest}
	\label{SDSS33_table}
	\begin{center}
	\begin{tabular}{lccr} % fifteen columns, alignment for each
		\hline\hline
		DESIGNATION & BPT & OBS & FEATURE  \\\hline
J012553.91-004811.1	&2-SF	&2	&c\\
J012835.33+022652.1	&SF+LINER	&1	&b\\
J014745.99+293329.9	&2-SF	&1	&a\\
J071657.26+410029.5	&2-SF	&1	&c\\
J071908.7+403341.7	&SF+LINER	&1	&b\\
J073239.24+255403.3	&SF+LINER	&1	&c\\
J073253.57+362541.6	&2-SF	&2	&c\\
J074437.9+191453.0	&SF+LINER	&2	&b\\
J080648.25+282516.7	&2-SF	&3	&b\\
J083021.04+364627.1	&SF+LINER	&4	&a\\
J085650.59+383929.4	&2-SF	&2	&b\\
J090413.93+522655.4	&2-SF	&2	&b\\
J090610.48+131224.9	&2-LINER	&4	&c\\
J091652.26+031724.5	&2-SF	&2	&b\\
J094020.59+320450.7	&2-SF	&1	&c\\
J102834.12+030128.6$^{*}$	&SF+Seyfert	&2	&b\\
J103600.37+013653.5$^{*}$	&2-Seyfert	&2	&b\\
J111117.46+192255.1	&2-SF	&2	&b\\
J112121.34+182243.0	&SF+LINER	&2	&a\\
J122105.25+113752.8	&2-SF	&2	&c\\
J123516.73+044215.6	&2-SF	&3	&c\\
J123600.85+100605.6	&2-SF	&2	&b\\
J124327.83+232811.4	&unknown	&2	&b\\
J125839.34+123842.0$^{*}$	&Seyfert+LINER	&2	&a\\
J132308.89+271437.9	&2-SF	&2	&b\\
J134003.64+193602.7	&2-SF	&4	&c\\
J134330.04+090614.6$^{*}$	&SF+Seyfert	&2	&a\\
J142522.3+141126.5$^{*}$	&Seyfert+LINER	&3	&c\\
J142722.83+200549.3$^{*}$	&Seyfert+LINER	&3	&c\\
J145731.02+070333.0	&2-SF	&2	&a\\
J174849.85+255526.6	&2-SF	&1	&a\\
J212713.49+010234.6	&2-SF	&3	&b\\
J224146.13+040131.9$^{*}$	&2-Seyfert	&1	&b\\
\hline\hline
\end{tabular}\\
   \end{center}
\textbf{Notes}. Column 1: LAMOST DR4 designation hhmmss.ss+ddmmss.s (J2000.0). Column 2: spectroscopic observation type of the target, 1 stands for the targets which are spectroscopically observed for the first time by LAMOST, 2 for targets observed by both LAMOST and SDSS, 3 for sources with primary and the possible secondary core observed both by SDSS, 4 for targets whose primary core is spectroscopically observed by LAMOST, while their counterpart core is observed by SDSS. Column 3: Feature analysis, ``a" stands for dual-cored feature, ``b" for close companion in late pairing stage, ``c" for more close companions, some of which display tidal tails indicating a recent interaction.
\end{table}

\subsection{Analysis of MaNGA 1-556501}
\label{sec:MaNGA analysis}

\begin{figure*}
\centering
\includegraphics[width=0.9\textwidth]{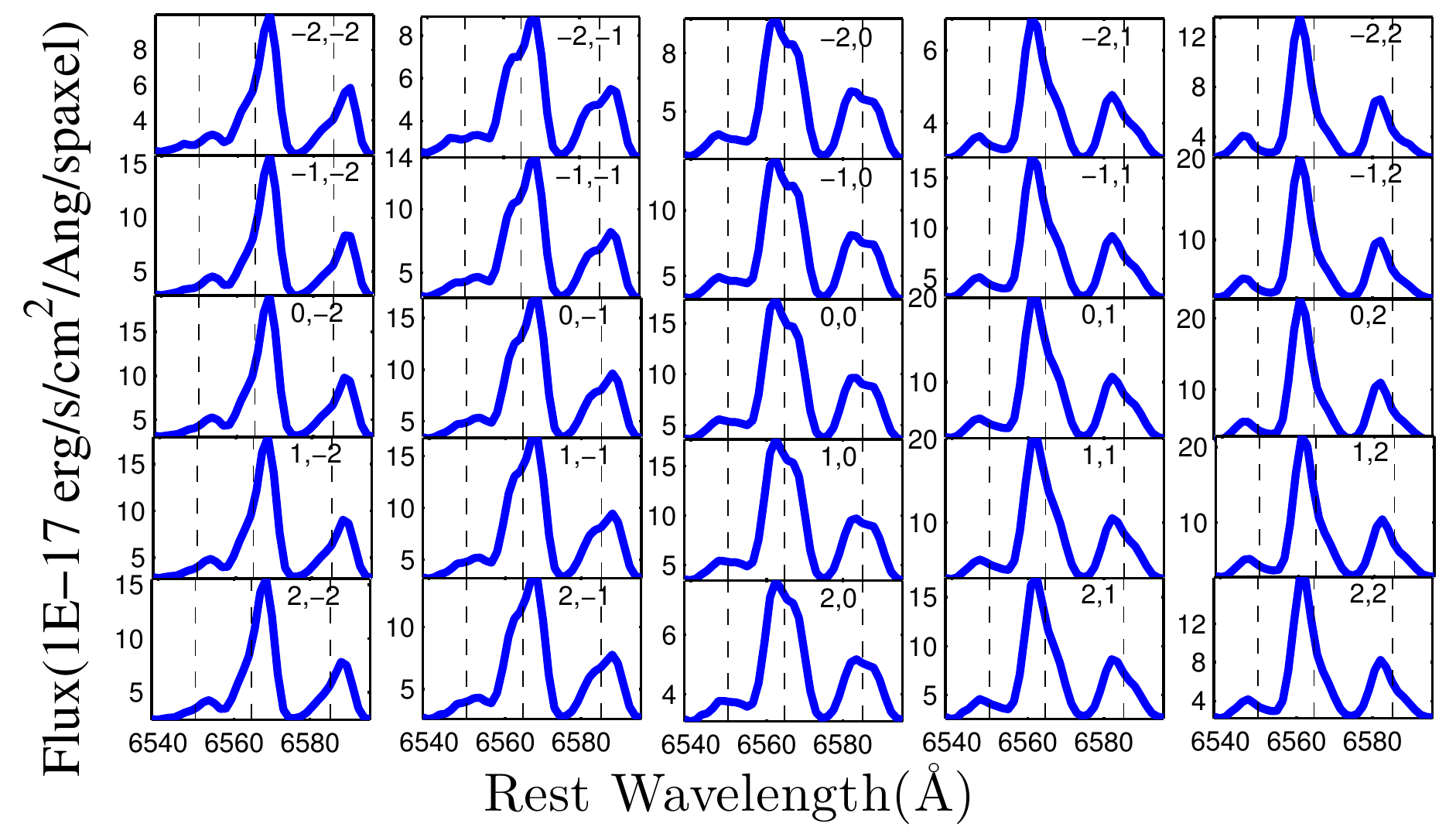}
\caption{MaNGA spectra for emission lines H$\alpha$ and [N~{\sc ii}]$\lambda\lambda6548,6584$ at the central 5 $\times$ 5 spaxels. The blue solid lines represent the extracted spectra from the MaNGA datacube. The vertical dashed lines indicate the wavelength centers of H$\alpha$ and [N~{\sc ii}]$\lambda\lambda6548,6584$ lines excepted from the redshift provided by MaNGA DRP.}
\label{ha_center_vary}
\end{figure*}

MaNGA uses IFU spectroscopy to measure spectra for hundreds of points within each galaxy, thus it can map the detailed composition and kinematic structure for each galaxy. Spectra for a target can be extracted from the MaNGA LOGCUBE, which has logarithmic wavelength sampling from log10($\lambda$/\AA)= 3.5589 to 4.0151 (NWAVE = 4563 spectral elements), and re-sampled 0.5 arcsec spaxels. Here we take the H$\alpha$ and emission lines as an instance, figure 5 provides the observed spectra for the central 5$\times$5 spaxels, covering a 2.5 arcsec $\times$ 2.5 arcsec region, the label (0,0) represents the innermost spaxel. As shown in figure \ref{ha_center_vary}, MaNGA can provide the flux and profile variations for emission lines within adjacent spatial pixels, the H$\alpha$ and [N~{\sc ii}] lines show distinct double-peaked features at the central spaxels, with flux rations between the two components varying spatially. 

\begin{figure*}
\centering
\includegraphics[width=0.31\textwidth]{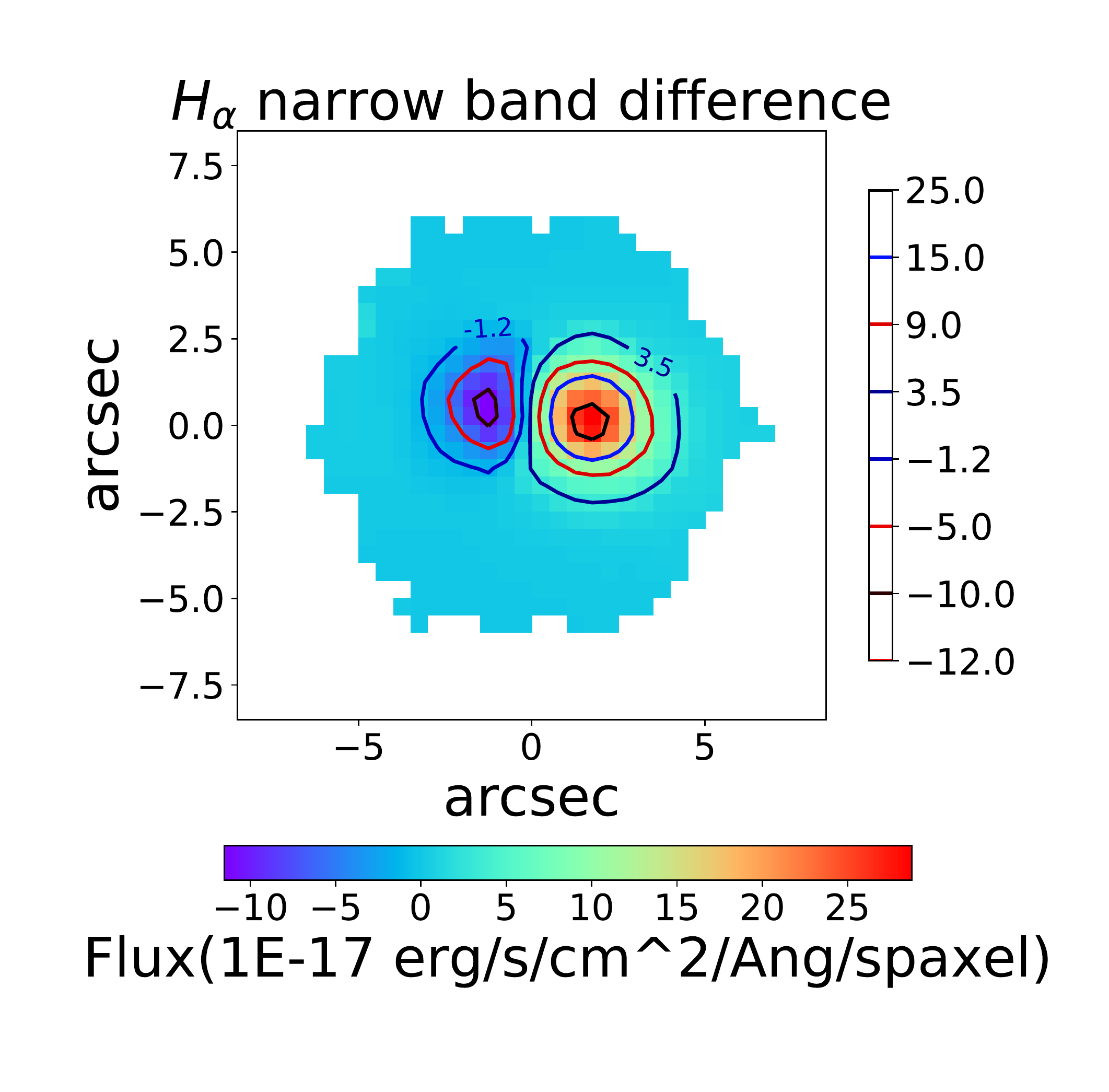}
\includegraphics[width=0.315\textwidth]{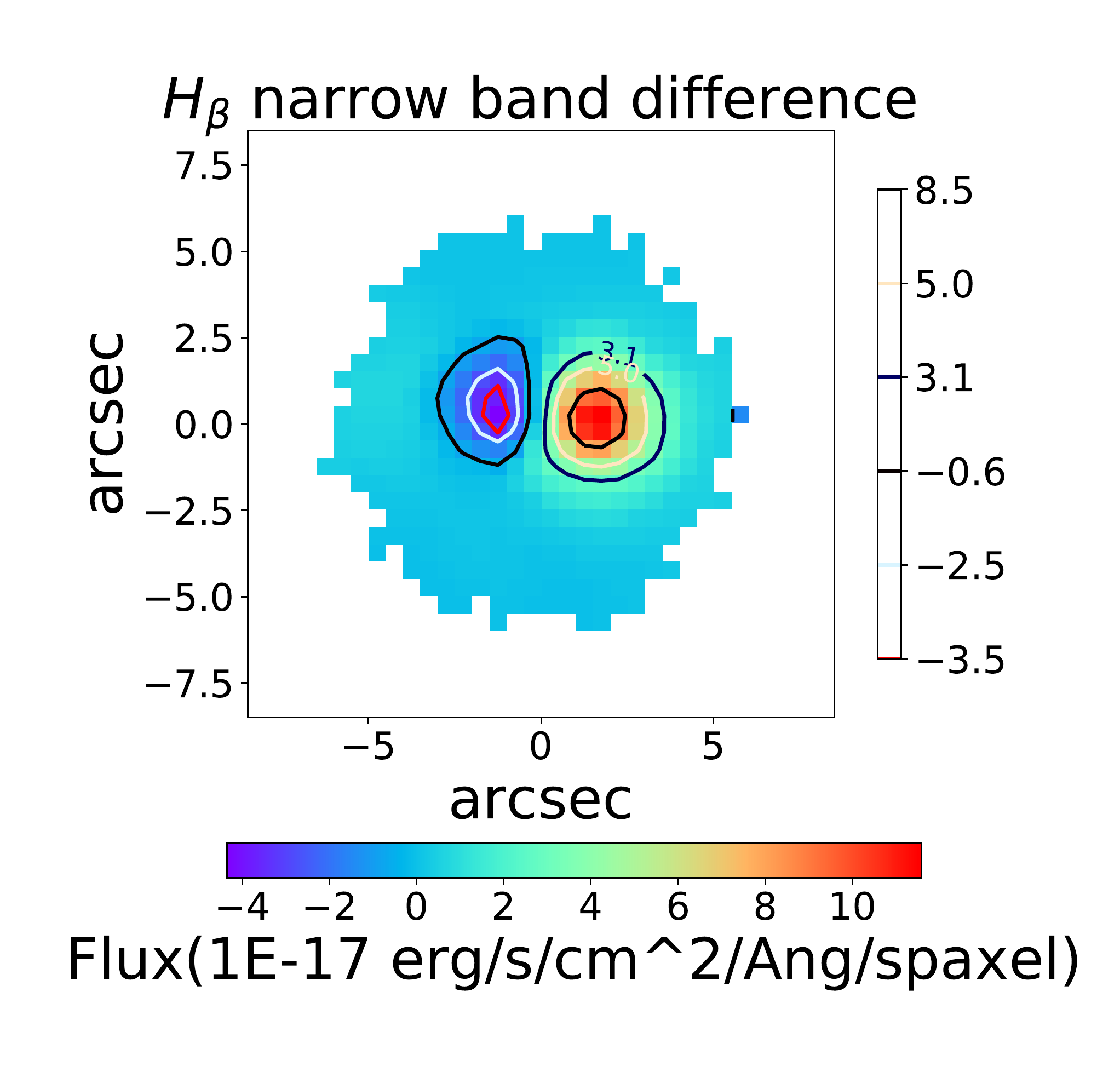}
\includegraphics[width=0.335\textwidth]{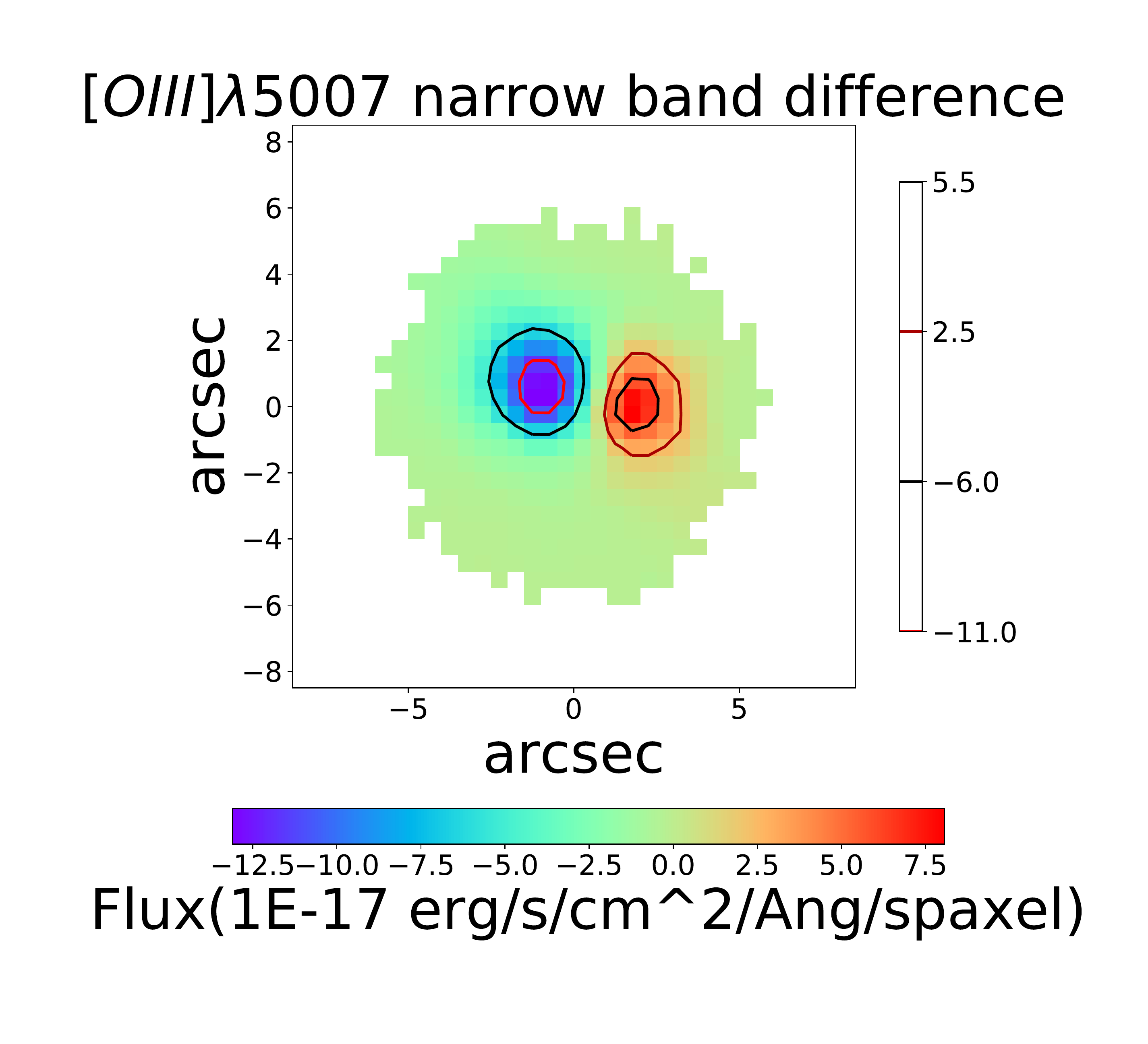}
\caption{The difference narrow band images of MaNGA 1-556501 for H$\alpha$, H$\beta$ and [O~{\sc iii}]$\lambda5007$ emission lines, plotted from MaNGA LOGCUBE. In each panel, one pixel represents a spaxel of 0.5 arcsec $\times$ 0.5 arcsec scale, flux contours are overlaid to reveal the values' distribution more clearly for NLRs of blueshifted and redshifted components, respectively. North is up and east is left.}
\label{narrow_band_diff}
\end{figure*}

We extract spectra from MaNGA LOGCUBE, and also decompose it with multi-gaussian models. From the fitting parameters, wavelength coverage for the blueshifted and redshifted components of each emission line can be determined, and thanks to the IFU datacube, the narrow band images of the two kinematic systems can be plotted respectively for each decomposed emission-line. To visualize the associated NLRs for each component more clearly, we resort to the narrow band image difference (here we subtract the redshifted system part from the blueshifted one), see figure \ref{narrow_band_diff}, which shows that regions of narrow lines for the kinematic components are spatially resolved, and for different lines, the core detected for each component is located proximately. The MaNGA IFU datacube enable us to obtain the flux centroid for each component. We employ the Gaussian guadrics fitting to model the narrow band difference and thus measure the relevant centroids. To get the more accurate core coordinates and better quantify the uncertainty of separation, we choose emission lines for 3 sets (H$\alpha$, H$\beta$ and [O~{\sc iii}]$\lambda$5007) of 9 central spaxels (27 emission line samples in total) from MaNGA LOGCUBE, and repeatedly calculate the flux centroids using the scheme mentioned above, finally we identify an average separation of 2.77 arcsec for the two components with a standard deviation of 0.043 arcsec. The two centers are separated by $\sim$1.6 kpc in physical projection at the distance of this target (z equals to 0.0271584 from MaNGA DRP). We also derive the nature of two kinematic systems by checking the ratios of the measured lines' flux in BPT diagram, the two components both fall into the composite category, the flux ratio for this target provided by the `EmissionlinesPort' table in SDSS CasJobs also reveals it a composite source. As it has been known that AGNs picked up by Kewley's criteria (theoretical separation lines) include only those that are dominated by nuclear activity, while AGNs selected via Kauffmann's criteria (empirical separation lines) would consist of AGN + starburst composite galaxies, this indicates the existence of AGN in this target. The resolved centers enable us to constrain the orientation and spatial positioning of the NLR. 

\citet{2013MNRAS.429.2594B} has found that only a minority of double-peaked narrow lines were directly induced by the relative motion of dual AGNs based on their hydrodynamic simulations of galaxy mergers. \citet{2015ApJ...813..103M} has critically investigated the nature of the ionizing sources of 18 optically identified double-peaked AGNs from SDSS, and pinpointed the origin of the double-peaked emission features for a sample of 18 optically identified double-peaked AGNs from SDSS, based on optical long-slit spectroscopy and high-resolution VLA multi-band observations in the range of 8-12 GHz. Among these, gas kinematics account for $\sim$ 75\% of the double-peaked narrow emission lines, while the remaining 10\% are ambiguous cases. Here we resort to a kinematic classification technique introduced by \citet{2016ApJ...832...67N} to discuss the scenarios that give rise to the double-peaked emission lines for this target. The quantitative classification scheme is created upon the 
origin determination of NLR for a complete sample of 71 Type II AGNs at {\it z} < 0.1 with double-peaked features from SDSS using the spatially resolved information from long-slit data alone, it classifies a spectrum into  ``Outflow-dominated" or ``Rotation-dominated" kinematics, and subdivides outflows as ``Outflow Composite" and ``Outflow" based on the number of best-fitting Gaussian components, and further classifies rotation-dominated spectra into subcategories of ``Rotation-dominated + Disturbance", ``Rotation-dominated + Obscuration" and ``Ambiguous". The important properties for kinematic nature identification of a galaxy are the velocity dispersion of either blueshifted or redshifted component from [O~{\sc iii}]$\lambda5007$ multi-gaussians fit, the line-of-sight velocity of [O~{\sc iii}]$\lambda5007$ derived from the one Gaussian fit of the spectra, the alignment of the [O~{\sc iii}]$\lambda5007$ line with the major axis of the galaxy, and the number of Gaussian components in spectral fitting. 

\begin{figure}
\centering
\includegraphics[width=0.4\textwidth]{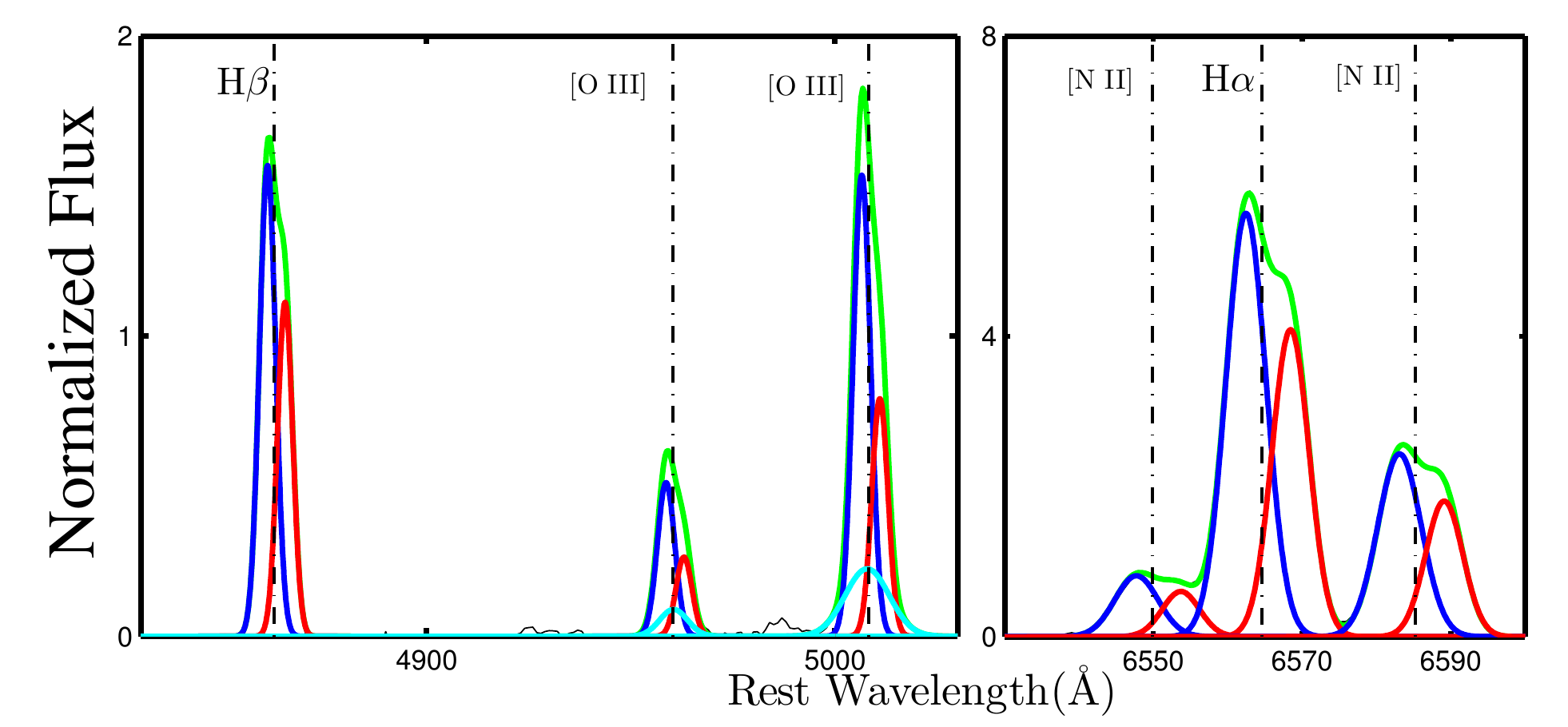}
\caption{Gaussian decomposition for the innermost spectra from MaNGA 1-556501 LOGCUBE. Two kinematic components plus [O~{\sc iii}] wings can nicely describe the H$\beta$ and [O~{\sc iii}]$\lambda\lambda4959,5007$ lines, while three double-gaussians successfully fit the H$\alpha$-[N~{\sc ii}]$\lambda\lambda6548,6584$ complex. The vertical dashed lines indicates the wavelength centers of H$\beta$, [O~{\sc iii}]$\lambda\lambda4959,5007$, H$\alpha$ and [N~{\sc ii}]$\lambda\lambda6548,6584$ lines excepted from the redshift (z = 0.0270934) measured from the host galaxy stellar absorption lines.}
\label{MaNGA_fitting}
\end{figure}

\begin{figure}
\centering
\includegraphics[width=0.4\textwidth]{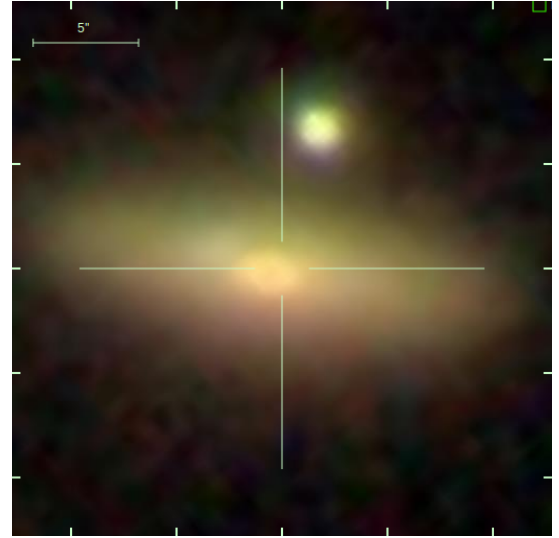}
\caption{SDSS color composite image for MaNGA 1-556501, and no apparent galaxies interacting can be seen from this image. The scale bar in the upper left corresponds to 5 arcsec. The image is centered to the coordinates from LAMOST designation, and for the orientation, north is up and east is left. }
\label{manga_sdss}
\end{figure}

\begin{figure}
\centering
\includegraphics[width=0.5\textwidth]{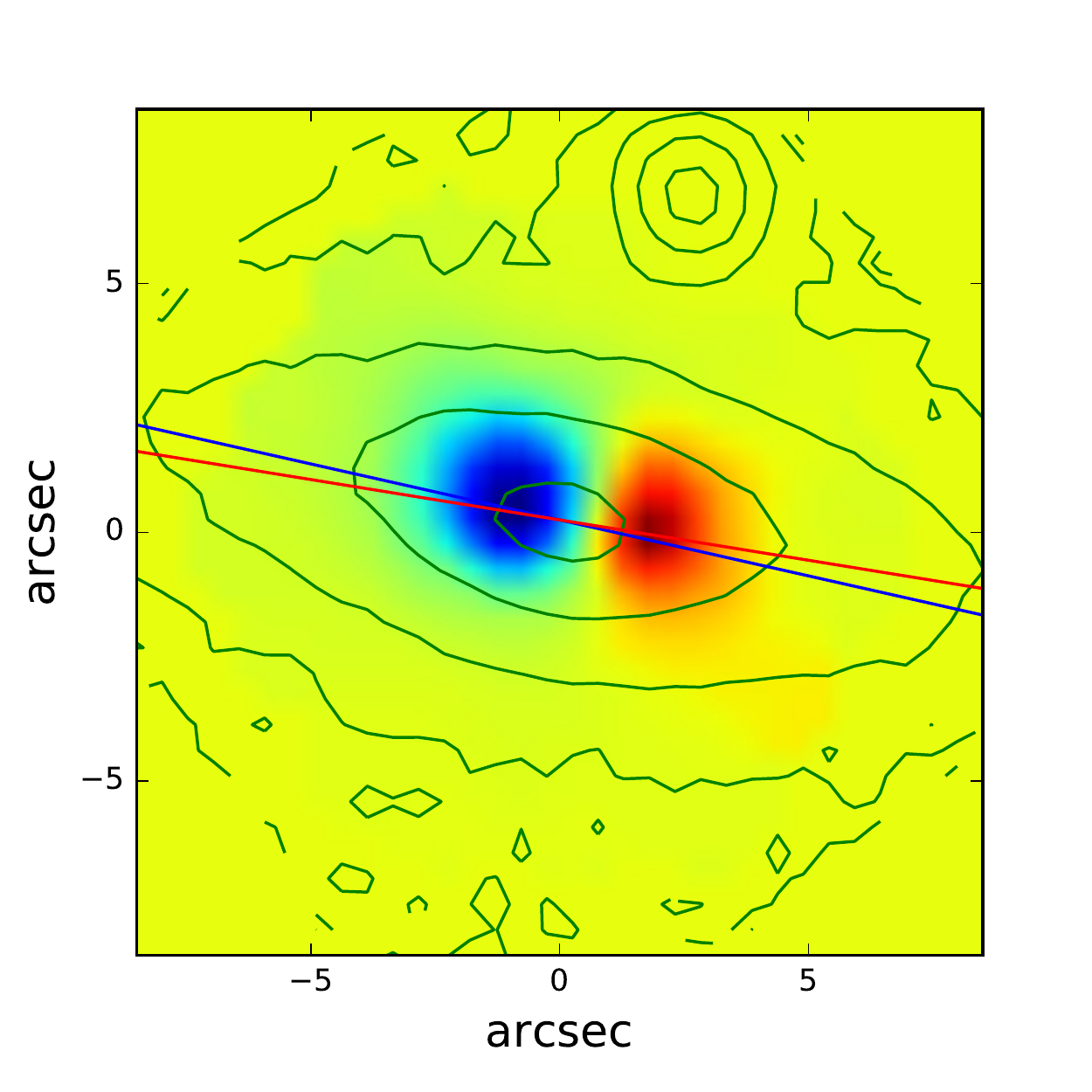}
\caption{GALFIT was used to decompose the target as a de Vaucouleurs core plus a exponential disk in g-band,the bulge is fitted with sersic n=4, and the galactic disk is fitted with sersic n=1, the contour show the best fitting result from GALFIT. The blue and red spots are overlaid to reveal the positions of the two kinematic components from gaussian guadrics fitting on [O~{\sc iii}]$\lambda5007$ narrow band difference. The blue line marks the photometric major axis of the target as measured from SDSS g-band photometry, while the red line indicates the position angle of
[O~{\sc iii}]$\lambda5007$ line as measured in this work.}
\label{img_fluxdiff_sdss}
\end{figure}

\begin{figure*}
\centering
\includegraphics[width=0.9\textwidth]{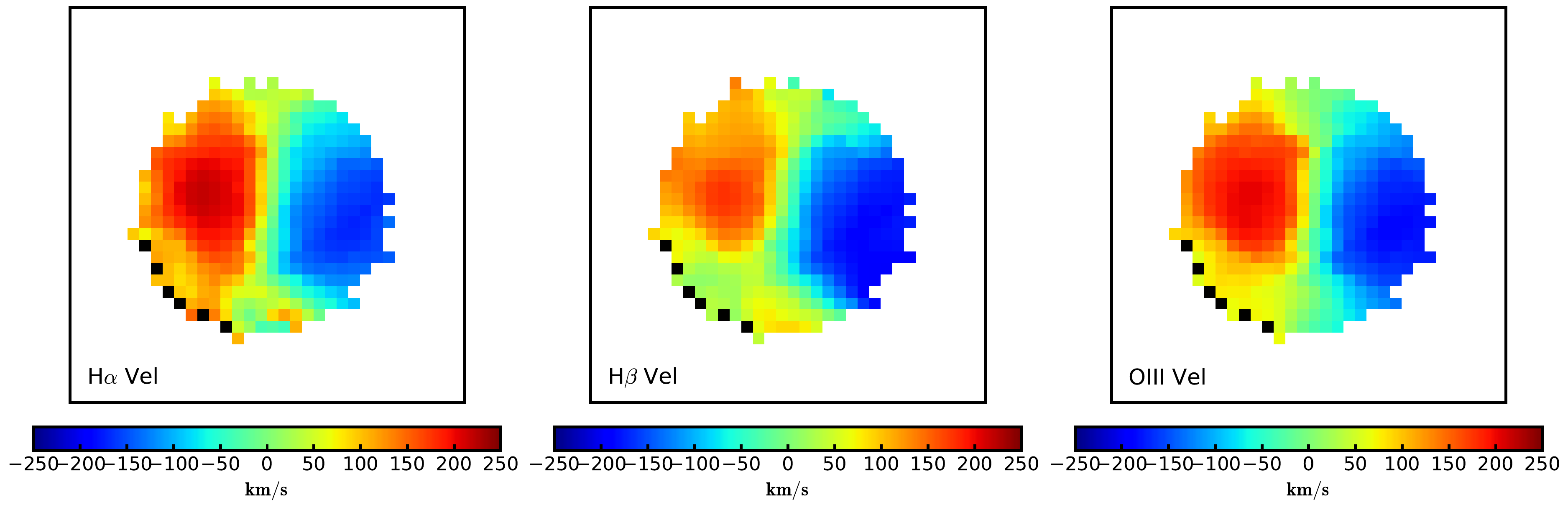}
\caption{Velocity maps of H$\alpha$, H$\beta$ and [O~{\sc iii}]$\lambda5007$, which are measured by gaussian profile fitting for each line separately, the stellar continuum is removed prior to fitting procedure by using the PPXF code \citep[see][]{2004PASP..116..138C} and the MIUSCAT simple stellar population models \citep[see][]{2012MNRAS.424..157V}.}
\label{vel_map}
\end{figure*}

We inspect the central 5 $\times$ 5 spaxels' spectra across the spatial extent of the galaxy from MaNGA LOGCUBE. For each spectrum, we fit the two sets of emission lines H$\beta$-[O~{\sc iii}] and H$\alpha$-[N~{\sc ii}] separately, and put consistent constraints on redshift and width within each lines set, then find that gaussians with similar widths and velocity offsets could remodel the two sets of emission lines perfectly. As one example displayed in figure \ref{MaNGA_fitting}, two kinematic components plus [O~{\sc iii}] wings can describe the H$\beta$-[O~{\sc iii}]$\lambda\lambda4959,5007$ lines well, for the innermost spectra, the velocity gap for H$\beta$-[O~{\sc iii}] set amounts to 264 km s$^{-1}$, with a value greater by 9 than it for the H$\alpha$-[N~{\sc ii}] complex, which is 273 km s$^{-1}$ (see table \ref{manga_table}). For each of these 25 spectra from different spatial positions, the blueshifted and redshifted components from multi-gaussians fitting are both narrow, with a maximum velocity dispersion of 203 and 177 km $\mathrm{s}^{-1}$ respectively, the single Gaussian fitting for each [O~{\sc iii}]$\lambda5007$ line also reveals that the maximum line-of-sight radial velocity is 139 km $\mathrm{s}^{-1}$, far below the 400 km $\mathrm{s}^{-1}$ limit, which is employed to differentiate ``Outflow-dominated" from ``Rotation-dominated" kinematics, as it has been analysed in \citet{2016ApJ...832...67N}, velocity constraints are the most powerful properties to differentiate ``Outflow-dominated" from ``Rotation-dominated" kinematics, since rotating structure should behave according to Keplerian physics, thus, placing limits on the radial velocity (V$_r <$ 400 km $\mathrm{s}^{-1}$) and velocity dispersion ($\sigma_1$ and $\sigma_2 < 500$ km $\mathrm{s}^{-1}$, \citet{2006agna.book.....O}). In this case, this double-peaked galaxy should not be classified as ``Outflow-dominated". The SDSS color composite image for MaNGA 1-556501 is displayed in figure \ref{manga_sdss}, and no apparent galaxies interacting can be seen from this image, which also strengths the possibility of ``Rotation-dominated" origin for this source.
We derive the position angle (PA$_{\rm gal}$) ($\sim$ $78^{\circ}$) of the 
major axis of the target from SDSS photometry, GALFIT 
\citep[see][]{2002AJ....124..266P} is performed to decompose this galaxy as a de Vaucouleurs core plus a exponential disk in g-band. The spatially distinct 
blueshifted and redshifted components from [O~{\sc iii}]$\lambda5007$ narrow band difference are overlaid on the GALFIT decomposition for this galaxy in figure \ref{img_fluxdiff_sdss}, where the blue line indicates the major axis of photometry image, and the red line indicates the PA of the [O~{\sc iii}]$\lambda5007$ line ($\sim$ $81^{\circ}$), which is denoted as PA$_{\mathrm{[O III]}}$. It can be clearly seen that the the two centers are aligned with the major axis of the galaxy (PA$_{\mathrm{[O III]}}$ $\sim$ PA$_{\rm gal}$). Alignment of the ionized gas with the stellar disk is a property of rotation-dominated galaxies, this situation has been discussed in \citet{2015ApJ...813..103M} and \citet{2016ApJ...832...67N}, where the PA$_{\mathrm{[O III]}}$ for the NLR were obtained from optical long-slit spectroscopy. Based on these properties, the most plausible explanation for this galaxy is that the double-peaked narrow emission lines are produced by rotational disk. Here we use a definition of asymmetry from \citet{2012ApJ...752...63S} to distinguish between ``Rotation-dominated + Disturbance" and ``Rotation-dominated + Obscuration" kinematics. The ratio of the redshifted component to the blueshifted flux component for each spectrum is derived from gaussian fitting, with only 2 out of the 25 spectrums displaying a value within $0.75 \leq F_r/F_b \leq 1.25$, which is used as a criteria in \citet{2012ApJ...752...63S} to classify their sample of ``equal-peaked" AGNs, here $F_r$ is the flux of the redder gaussian component and $F_b$ is the flux of the bluer component. Although their quantitative classifiction involves only double-gaussians fitting, while for this target, the [O~{\sc iii}]$\lambda5007$ is better fitted by two kinematic components plus a low flux blue wing, we can see from figure \ref{MaNGA_fitting} that the wing component contributes little to the flux ratio between bluer and redder components, thus the asymmetric value employed here is still reliable. Considering the high degree of asymmetry of the [O~{\sc iii}]$\lambda5007$, we further classify this target as a rotation dominated object with asymmetric double-peaked features due to disturbances from nuclear bars, spiral arms, or a disturbed dual AGNs \citep[e.g.][]{1997MNRAS.292..349S,2009ApJ...702..114D,2013MNRAS.429.2594B,2016ApJ...832...67N}. This model can also nicely explain the consistency of velocity offsets, velocity dispersions and relative asymmetric profiles between H$\alpha$ and [O~{\sc iii}]$\lambda5007$ emission lines, since H$\alpha$ traces both the NLR and stellar kinematics, and in the case of a rotation-dominated NLR, the ionized gas kinematics should be identical to the stellar kinematics of stars in the disk, which has also been verified in \citet{2016ApJ...832...67N}. The velocity maps of H$\alpha$, H$\beta$ and [O~{\sc iii}]$\lambda5007$ are provided in figure \ref{vel_map}, which show consistent velocity fields between these emission lines, and provide strong kinematic evidence for our analysis on this object.

\begin{table*}
	%\centering
	\caption{MaNGA 1-556501 fit results}
	\label{manga_table}
	\begin{center}
	\begin{tabular}{lcccccr} % fifteen columns, alignment for each
		\hline\hline
		*&H$\beta$&	[O~{\sc iii}]$\lambda$4959&[O~{\sc iii}]$\lambda$5007&H$\alpha$&	[N~{\sc ii}]$\lambda$6548&[N~{\sc ii}]$\lambda$6584 \\ \hline
		\multicolumn{7}{l}{Blue System} \\ \hline
		$\Delta V^{a}$&-99&	-99$^{c}$&	-99$^{c}$&	-97&	-97$^{c}$&	-97$^{c}$\\
		FWHM$^{b}$&281&	296&	296$^{c}$&	283&	307&	307$^{c}$ \\ \hline
		\multicolumn{7}{l}{Red System} \\ \hline
		$\Delta V^{a}$ &  165& 165$^{c}$ & 165$^{c}$   & 176 &  176$^{c}$  & 176$^{c}$\\
		FWHM$^{b}$&264   &	  272& 272$^{c}$ & 259            & 262    &   262$^{c}$ \\ \hline\hline
 \end{tabular}
 \end{center}
 
     \textbf{Notes}.
     $^{\rm a}$ In units of km s$^{-1}$.
     $^{\rm b}$ In units of km s$^{-1}$. Corrected for the instrumental resolution. 
     $^{\rm c}$fixed.
 
\end{table*}

\section{SUMMARY}

Our search is motivated by the prospect of discovering more merging galaxies and dual AGN candidates, which are interesting targets for detailed analysis. We set up a large sample, composed of 188 objects with double-peaked narrow lines and 137 targets displaying asymmetric profiles, based on the LAMOST DR4 dataset. In addition, we provide 33 targets therein revealing meaningful features in the optical images. Among these, there are 7 dual/offset AGN candidates showing signatures of interactions, which are the most promising dual/offset AGN candidates found in our study, and the best candidates for follow-up analysis.
 
A target is further stressed based on the the spatial and kinematic decompositions of the narrow emission lines from both the LAMOST and MaNGA IFU spectroscopy. Thanks to the high-spatial-resolution LOGCUBE provided by IFU, we could track the emission line profiles across the spatial extent of the target and further constrain the origin responsible for the  double peaks in this galaxy. Considering that its ionized gas emission is spatially coincident with the major axis of the galaxy, and the kinematic decomposition of this target also reveals it as one object with narrow kinematic components, low line-of-sight radial velocity and high-degree asymmetric [O~{\sc iii}] profile, we conclude that the most promising interpretation for the origin of double peaks in LAMOST J074810.95+281349.2 is a `Rotation Dominated + Disturbance' model.

\section*{ACKNOWLEDGEMENTS}

This work is sponsored by the Funds of the National Natural Science Foundation of China (Grant No.11603042, 11673081). We thank Professors Jianrong Shi and Xiaoyan Chen for providing valuable advice on this work. 
 
This work has made use of data products from the Large Sky Area Multi-Object Fibre Spectroscopic Telescope (LAMOST), Sloan Digital Sky Survey (SDSS), Mapping Nearby Galaxies at the Apache Point Observatory (MaNGA) survey. Thanks for their tremendous efforts on the surveying work.

\end{document}